\documentclass[letterpaper,12pt,margin=1in]{article} 
\usepackage{authblk}
\usepackage{fullpage}
\usepackage{graphicx}
\usepackage{epstopdf}
\usepackage{txfonts}
\usepackage{subfigure}
\usepackage{footnote}
\usepackage{lineno}
\usepackage{lscape}
\usepackage[authoryear,round]{natbib}

\begin{document}

\title{Numerical simulations of ICME-ICME interactions}
\author[1]{Tatiana Niembro}
\author[2]{Alejandro Lara}
\author[3]{Ricardo F. Gonz\'alez}
\author[4]{J. Cant\'o}
\affil[1]{Posgrado en Ciencias de la Tierra, Universidad Nacional Aut\'onoma de M\'exico}
\affil[2]{Instituto de Geof\'isica, Universidad Nacional Aut\'onoma de M\'exico, Ciudad Universitaria, CDMX, 04510, M\'exico}
\affil[3]{Instituto de Radioastronom\'ia y Astrof\'isica, Universidad Nacional Aut\'onoma de M\'exico, Apdo. Postal 3-72, Morelia, Michoac\'an, 58089, M\'exico}
\affil[4]{Instituto de Astronom\'ia, Universidad Nacional Aut\'onoma de M\'exico, Ciudad Universitaria Apdo. Postal 70-264, CDMX, 04510, M\'exico}

\maketitle 
	 
\abstract
{We present hydrodynamical simulations of interacting Coronal Mass Ejections in the Interplanetary medium (ICMEs). In these events, two consecutive CMEs are launched from the Sun in similar directions within an interval of time of a few hours. In our numerical model, we assume that the ambient solar wind is characterized by its velocity and mass-loss rate. Then, the CMEs are generated when the flow velocity and mass-loss rate suddenly change, with respect to the ambient solar wind conditions during two intervals of time, which correspond to the durations of the CMEs. After their interaction, a merged region is formed and evolve as a single structure into the interplanetary medium. In this work, we are interested in the general morphology of this merged region, which depends on the initial parameters of the ambient solar wind and each of the CMEs involved. In order to understand this morphology, we have performed a parametric study in which we characterize the effects of the initial parameters variations on the density and velocity profiles at 1 AU, using as reference the well-documented event of July 25th, 2004. Based on this parametrization we were able to reproduce with a high accuracy the observed profiles. Then, we apply the parametrization results to the interaction events of May 23, 2010; August 1, 2010; and November 9, 2012. With this approach and using values for the input parameters within the CME observational errors, our simulated profiles reproduce the main features observed at 1 AU. Even though we do not take into account the magnetic field, our models give a physical insight into the propagation and interaction of ICMEs.}   


\section{Introduction}

CMEs are powerful solar eruptions that release huge amount of mass into the interplanetary medium (IPM). Their masses can be as large as 10$^{15}$-10$^{16}$ g moving outwards at speeds ranging from a few hundreds to thousands kilometers per second (Hundhausen, 1999). Interaction of these eruptions with the solar wind lead to the formation of shock waves that travel through the IPM (Pizzo, 1985). Several authors have investigated the CME propagation through the inner heliosphere (Sun-to-Earth), which is a fundamental issue in space weather forecasting (e.g. Gopalswamy, 2001; Vr$\check{\mbox{s}}$nak,  2001; Borgazzi et al. 2009; Liu et al. 2013). On the other hand, few authors have addressed the dynamics of interacting events (Liu et al. 2012; Temmer et al. 2012). In such cases, two consecutive CMEs launched from the Sun in similar directions collide and the resulting merged structure interacts with the ambient solar wind (Burlaga et al. 2002; 2003). Based on the high rate of CME production (Yashiro et al. 2004; Gopalswamy et al. 2016), it is possible that most of the ICMEs observed at 1 AU may be formed by the interaction of two or more CMEs. This may be the main cause the failure in the models that attempt to predict the travel time of a single CME. Thus, it is important to study both the CME propagation through the IPM and the dynamics of the ICME-ICME interaction events. The physics of these phenomena is not yet well understood, and hence, it is still one of the goals for space research.

The detection of CMEs near the Sun can be achieved by coronagraphic observations, while their counterparts ICMEs can be detected by multi-spacecraft {\it in situ} measurements. Consequently, the identification of CME-ICME corresponding pairs is difficult (Lara et al. 2006), just in a very few cases, it has been possible to track ICMEs from the Sun to the Earth by using the STEREO heliospheric imager (see, for instance, Harrison et al. 2012; Lugaz et al. 2012; Temmer et al. 2012) but the analysis of these images is not enough to understand the physics behind their evolution. From a theoretical point of view, both, analytical and numerical models have been developed for a better understanding of the ICMEs dynamical evolution, propagation and possible interaction among these structures. Both analytic models (e.g. Vr$\check{\mbox{s}}$nak 2001; Borgazzi et al. 2009; Vr$\check{\mbox{s}}$nak et al. 2013) and numerical simulations (e.g. Xiong et al. 2007; Shen et al. 2011; 2013; Lugaz et al. 2013) describe the evolution and propagation of a single ICME by considering that the ICME dynamics is dominated by the aerodynamic drag. In general terms, the ICME kinematics depends on the ICME and the ambient solar wind conditions (speed and density) and a dimensionless drag coefficient. However, the number of free parameters involved causes uncertainties in forecasting the ICME arrival. Even more, these uncertainties increase when the ICME-ICME interaction takes place.  Currently, the prediction errors ranges from $\pm$10 to 24 hours. The numerical models have good results about the prediction of the arrival magnitude of the velocity and density of ICMEs, although, their performance is relatively poor in terms of the time profile morphology of the {\it in-situ} parameters (see for instance, Lugaz et al. 2008; Temmer et al. 2011; Manchester et al. 2014). In this work we address this issue.

Recently, Niembro et al. (2015) have performed an analytic study of ICMEs interactions. These authors applied the formalism developed by Cant\'o et al. (2005) for velocity fluctuations in the solar wind, in order to study the dynamics of two consecutive CMEs launched in the same direction from the Sun. The collision yields a merged region that contains material expelled during both eruptions and propagates afterwards as a single structure. The time and the distance of the collision, as well as the arrival time at 1 AU of the merged region is predicted for a set of well documented interaction events (January 24, 2007; May 23, 2010; August 1, 2010; and November 9, 2012) with a error of less than 2 h, but the {\it in-situ} parameter time profiles can not be predicted with this approach. To address this issue, in this work, we have performed hydrodynamical simulations for a single ICME and for the ICME - ICME interaction events. We perform a parametric study of a single ICME to analyze which are the most relevant quantities for the time profile morphology, and then to obtain the best numerical fit of the density and velocity profiles. 

This paper is organized as follows. In $\S$2, we describe the numerical models. In $\S$3, we present the parametric study for the dynamics of a single CME as function of the input parameters. The numerical simulations of ICME-ICME interaction events, and the discussion of the results are presented in $\S$4. Finally, in $\S$5 we give our conclusions. 


\section{The numerical models}

We investigate through numerical simulations the propagation of a single CME into the solar wind, as well as ICME-ICME interaction events. We focus on the ICME dynamics (macroscopic scale), assuming that the kinetic energy of the plasma is much higher and therefore dominates over the magnetic field energy (see, for instance, Borgazzi et al. 2009).

In our models, we assume that, a CME, is a perturbation in which both, the ejection velocity and the mass-loss rate suddenly increase, with respect to the pre-eruptive wind parameters, by a constant factor (different for each parameter) and during a finite interval of time to then resume the solar wind conditions again (see Figure 2 in Niembro et al. 2015). This step function forms an outgoing two-shock structure, with an inner shock that decelerates the fast ICME material, and an outer shock that accelerates the solar wind (see Appendix A). It has been shown analytically by Cant\'o et al. (2005) that two different stages in the dynamical evolution of this structure are expected. Initially, the shocked layer propagates with constant velocity, which is intermediate between the background wind speed and the velocity of the material ejected during the eruption. In this first stage, the shocked material forms a dense shell which grows in mass with time. Once all the fast material expelled during the CME incorporates into the shell, the inner shock disappears while the outer shock continues accreting the ambient solar wind, giving rise to the second stage of its dynamical evolution in which the shell decelerates asymptotically approaching the solar wind speed. 

In the particular case of two consecutive CMEs launched in similar directions into the IPM, the corresponding ICMEs may collide before their arrival to the Earth, basically depending on the initial parameters of the individual events and the interval of time between them. Niembro et al. (2015) developed an analytical model of this interaction showing that the leading eruption (ICME$_1$) evolves as described above for the case of a single eruptive event, that is, the shocked layer suffers two different stages in its dynamics: an initial constant velocity phase and a final deceleration phase. Nevertheless, the trailing eruption (ICME$_2$) leads to the formation of a second shocked layer which evolves as follows: an initial constant velocity phase in which this layer is fed by both the ICME$_2$ and the pre-eruptive solar wind; a second acceleration or deceleration phase, depending in which material, from the pre-eruptive solar wind or from the ICME$_2$ incorporates first to the shocked layer (compression region), respectively; and a last phase of constant velocity after the ending of the accelerated/decelerated phase. Adopting the flow parameters of a set of well observed ICME-ICME interaction events, Niembro et al. (2015) showed that the collision may occur in either stage of evolution of each ICME. 

At 1 AU, an ICME is characterized by the arrival of a shock front driven by a compression region which is followed by a low density structure (the CME) that corresponds to a rarefaction zone. In our model, the shock structures correspond to the compression regions formed due to the interaction between the ICMEs with the solar wind. In the present paper, we extend the models developed by Cant\'o et al. (2005) and Niembro et al. (2015) to the more realistic case of numerical simulations. Firstly, we consider the propagation of a single ICME, and present the results of a parametric study developed for investigating the dynamical evolution of the shocked layer (compression region) and the rarefaction zone in terms of the initial conditions of both the solar wind and the eruptive event. Secondly, we use these results to reproduce the 1 AU observations of ICME-ICME interaction events.

We select those events in which the ICMEs propagation axes are very close to the Sun-Earth direction with high speeds and masses which allow us to neglect deflection and rotation effects due to the magnetic field (Kay et al. 2015). This means, that the evolution of the compression regions are dominated by the kinetic energy of the ICMEs (Siscoe et al. 2008) and are independent of the internal magnetic field of the ICMEs (Riley et al. 2004; Owens et al. 2005). The {\it in situ} observatory is localized in the nose path of the ICME, allowing us to neglect geometry and projection effects (Schwenn et al. 2005). Deviations of these assumptions are considered limitations of our 2D model.

The numerical simulations have been performed using a 2D hydrodynamic version of the adaptive grid code YGUAZ\'U-A, originally developed by Raga et al. (2000) and modified by Gonz\'alez et al. (2004a,b; 2010). The original code integrates the hydrodynamic equations for the atomic/ionic species HI, HII, HeI, HeII, HeIII. The adopted abundances by number are H=0.9 and He= 0.1. The flux-vector splitting algorithm of Van Leer (1982) is employed. The simulations were computed on a five-level binary adaptive grid with a maximum resolution of 1.465 $\times$ 10$^{10}$ cm, corresponding to 1024 $\times$ 1024 grid points extending over a computational domain of ($1.5 \times 10^{13}$ cm) $\times$ ($1.5 \times 10^{13}$ cm) and time steps of 15 min. This code allowed us to perform numerical simulations with better spatial and temporal resolution than those models developed with the ENLIL code of our study cases which were done with a time step of 8 min and a grid resolution of 256 x 30 x 90 (Falkenberg et al. 2010).

The computational domain is filled from an injection radius $R$ up to 1 AU by an isotropic flow with typical solar wind conditions, speed $v_{\mbox{\tiny SW}}$ and mass-loss rate $\dot m_{\mbox{\tiny SW}}$.  It is well known that the Sun has a strong magnetic field at its surface. Because the solar wind is completely ionized, it has a high electrical conductivity and thus, the solar magnetic field is frozen into the solar wind. Close to the Sun the solar magnetic field is very strong and dominates the dynamics of the solar wind. However, at larger distances, the solar magnetic field is quite weak (the radial component decreases as the inverse of the distance squared) and thus, not far from the Sun the dynamical influence of the solar magnetic field on the dynamics of the solar wind becomes negligible . The injection radius was estimated as the distance at which the role of the magnetic field changes by comparing the kinetic energy of the flow with the magnetic energy and find that the Alfvenic critical point is of only 20 R$_\odot$ (Mihalas, 1978). Then, the first eruption CME$_1$ is launched during an interval of time $\Delta t_1$ within a solid angle $\Omega_1$ with an ejection speed $v_{1}\,(> v_{\mbox{\tiny SW}})$ and a mass-loss rate $\dot m_{1}$. Afterwards, the solar wind resumes until the second eruption CME$_2$ is expelled (with $v_2\,> v_{1}$, $\dot m_{2}$, $\Omega_2$, during $\Delta t_2$). After the CME$_2$, the solar wind conditions again resumes.


\section{Parametric study of the dynamics of a single CME}

In order to explain the time-dependent profiles of the observed {\it in-situ} physical properties of the solar wind, we have performed a parametric study of the dynamics of a single CME as a function of the injection parameters. Taking into account that the {\it in-situ} observed morphology of a fast ICME is characterized by a shock front, a compression region and a rarefied zone, in this section, we present a comparison between observations of the July 25, 2004 event with numerical models computed by the YGUAZ\'U-A code. 

Vr$\check{\mbox{s}}$nak et al. (2010) studied this event using different drag based models to investigate the CME kinematics assuming that the velocity, the mass loss rate and the angular width of the eruption, as well as the solar wind conditions change over time. These authors focused on the CME transit-time, density, and speed at 1 AU. To compare our numerical predictions with their results, we have performed three distinct simulations using the mean and extremes (minimum and maximum) values of the angular width and solar wind density adopted by these authors.

In the Sun-Earth event of July 25, 2004, a CME was detected with a speed $v =$ 1330 km s$^{-1}$, in which a total mass $m =$ 1.1 $\times$ 10$^{16}$g was expelled during an interval of time $\Delta t =$ 2.0 h within a solid angle $\Omega/4\pi$ ranging from  0.07 (which corresponds to an angular width of $\phi =$ 1.1 rad) to 0.36 ($\phi =$ 2.6 rad). Here, we adopt for the solar wind a terminal velocity $v_{\mbox{\tiny SW}} =$ 650 km s$^{-1}$, and a number density $n=$ 0.5 - 2.5 cm$^{-3}$ at 1 AU.  The corresponding ICME was observed by the WIND spacecraft at a transit time $TT_{obs}=$31.5 h and a velocity of $v_{obs}=$1033 km s$^{-1}$. All these parameters are consistent with those reported by Vr$\check{\mbox{s}}$nak et al. (2010).

\begin{figure} [!h]
\centering
\includegraphics[scale=0.68, angle=90]{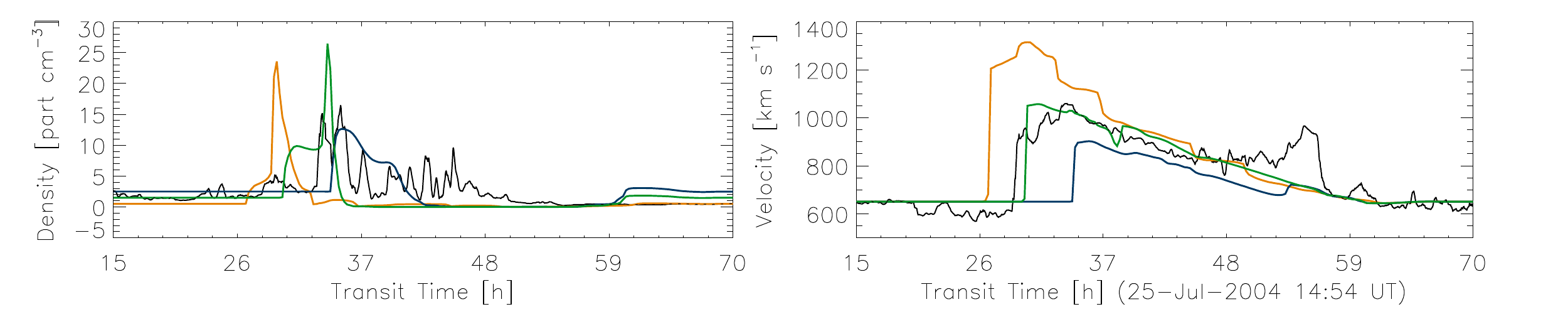} 
\caption{Density ({\it left panel}) and velocity ({\it right panel}) profiles as function of time predicted for the models listed in Table 1. The model MEDIUM, which is the best fit to observations, is shown in {\it green}. The profiles in {\it orange} correspond to model MAX, while the profiles in {\it blue} are calculated from  model MIN. The {\it in situ} data obtained by WIND are shown in black with a smoothed time of 15 min, which coincides with the time resolution of our simulations.} 
\label{Fig:f1}
\end{figure}

Table 1 shows our numerical results for the MIN, MEDIUM, and MAX models, which correspond to the minimum, mean and maximum values of $\phi$ and $n$, reported by Vr$\check{\mbox{s}}$nak et al. (2010). The first column shows the corresponding CME angular widths and the second the solar wind densities at 1 AU. Third, fourth, fifth and sixth columns are the transit time TT, the difference between the observed transit time and the computed value $\Delta TT= TT_{obs} - TT$, the predicted velocity $v$, and the difference between the observed and calculated velocities $\Delta v = v_{obs} - v$, respectively.

\begin{table}[!h]
\caption{Comparison between the predicted and observed arrival times and speeds at 1 AU.}
\begin{center}
\begin{tabular}{c c c c c c c}
\hline
\hline
 & $\phi$ & $n$ & $TT$ & $\Delta TT$ \footnotemark[1] & $v$ & $\Delta$v \footnotemark[2] \\
  &  rad  & part cm$^{-3}$ & h & h & km s$^{-1}$ & km s$^{-1}$ \\ 
\hline
MIN & 2.6 & 2.5 & 36.25 & -3.65 & 888 & 112 \\
MEDIUM &1.8 & 1.5 & 31.5 & 1.1 &  1033 & -33 \\
MAX & 1.1 & 0.5 & 30.25 & 2.35 & 1319 & -319 \\
 \hline
\end{tabular}
\end{center}
\footnote{} $\Delta TT = TT_{obs} - TT$, being $TT_{obs}$ the observed transit time and $TT$ the calculated value.  \\
\footnote{} $\Delta v = v_{obs}$- $v$, where $v_{obs}$ and $v$ are the observed and calculated speeds at 1 AU, respectively. \\
\end{table}

In model MEDIUM, we forecast that the ICME delays 1.1 h with respect to the observed arrival time, reaching the Earth with a velocity that differs only by $\Delta v=$-33 km s$^{-1}$ with respect to the observed 1 AU speed. The largest transit time, $TT$=36.25 h, is obtained with model MIN, which implies a  relative difference of $\Delta TT =$-3.65 h. As concerns the arrival velocity, we predict that the major deviation from the observed speed is $\Delta v =-$319 km s$^{-1}$ predicted by model MAX. 

Figure \ref{Fig:f1} shows the density and velocity profiles predicted by our numerical simulations. Models MIN, MEDIUM, and MAX are depicted with {\it blue}, {\it green} and {\it orange lines}, respectively. A comparison with the {\it in situ} data, obtained by WIND ({\it black line}), is presented. As it can be seen in the figure, the shape of the predicted profiles depends on the set of initial parameters of both the solar wind (mass-loss rate and ejection velocity) and the CME (speed, total mass, angular width, and duration). Based on these changes of the computed profiles, to further investigate this dependence, we carried out several numerical simulations varying the initial conditions of the outflows. This parametric study may help to determine the most relevant variables for the ICME forecasting. 

In order to get an insight of the studied phenomena, we have varied the initial parameters of the CME and study their impact on the predicted physical properties of the corresponding ICME at 1 AU. We have adopted the July 25, 2004 event (previously described as the model MEDIUM) as reference, and then, we vary one of the injection parameters while the others are fixed. In the next subsections, we present the results obtained by computing different models changing both the CME injection parameters ($\S$3.1), and the solar wind conditions before and after the eruption ($\S$3.2).


\subsection{Numerical models assuming different parameters of the CME}

The three relevant parameters of the CME were modified in our analysis: the ejection velocity $v$, the expelled mass $m$, and the duration of the eruption $\Delta t$. In this case the solar wind parameters are fixed with a speed $v_{\mbox{\tiny SW}}=$650 km s$^{-1}$ and a mass-loss rate $\dot m_{\mbox{\tiny SW}}=$9.27 $\times$ 10$^{-15}$ M$_{\odot}$ y$^{-1}$. As reference values for the parameters, we use the model MEDIUM described in $\S$2, that is, $v =$1330 km s$^{-1}$, $m=$1.1 $\times$ 10$^{16}$ g, and $\Delta t =$2.0 h. In Figure 2, we present the predicted density and velocity profiles at 1 AU ({\it left} and {\it right panels}, respectively) adopting different parameters of the CME: the expelled mass ({\it top panels}), the duration of the CME ({\it middle panels}), and the ejection velocity ({\it bottom panels}). The color bars at the right side of the figure shows the corresponding values for each color line. The {\it black lines} represents the reference density and velocity profiles obtained from the model MEDIUM.

We have performed ten simulations varying $\pm$15\%, $\pm$30\%, $\pm$45\%, $\pm$60\%, and $\pm$75\% the reference value of $m$. The resulting density and velocity profiles are shown in {\it panels} (a) and (b) of Figure 2, respectively. The profile in {\it dark-red} corresponds to the lowest mass (2.75 $\times$ 10$^{15}$ g), while the profile in {\it light-orange} corresponds to the highest mass (1.925 $\times$ 10$^{16}$ g).

We can see from density profiles that, as the expelled CME mass decreases, the ICME structure (shock front plus compression region) last for a longer time, and the transit time is longer. In addition, the density profiles show that the compression region morphology smoothens (the two peaks vanished) and the rarefaction is less deep. On the other hand, the velocity profiles predicts stronger arrival shocks (higher velocity jumps) for higher mass eruptions. Also, it can be observed in each profile, a second peak velocity that may be related with a contact discontinuity evolving into the rarefied zone. It is noticeable that this second jump is more perceptible for models with high mass eruptions.

Furthermore, we carried out numerical simulations in which the duration of the eruption was changed. In the {\it middle panels} of Fig. 2, we present the density and velocity profiles ({\it left} and {\it right}, respectively) which are predicted by ten different simulations. The adopted time variation is of $\pm$20 min, ranging from the minimum duration of 20 min ({\it dark blue}) to the maximum duration of 220 min ({\it light blue}). It can be seen from the figure that the adopted variations of the CME duration result in arrival times of the shock structure at 1 AU that differ less than one hour among the models. We also note that the main features of the density profiles are very similar to that obtained from the reference model MEDIUM ({\it black line}), while the velocity profiles show a second maximum more distant behind the leading shock front as the duration of the eruption increases.

Finally, we computed ten more simulations varying the injection velocity of the CME by $\pm$10\%, $\pm$20\%, $\pm$30\%, $\pm$40\%, and $\pm$50\%, with respect to the model MEDIUM. In Figure 2 ({\it bottom panels}), the predicted density and velocity profiles are presented. The {\it darkest green} represents a model with an injection velocity of 665 km s$^{-1}$, while the {\it lighter one} corresponds to a model with an initial speed of 1995 km s$^{-1}$. As expected, it is shown that the arrival time at 1 AU is shorter as the CME velocity increases. In addition, it can be seen from the density profiles that the detection of the compression region, and the corresponding rarefaction zone as well, last for longer times in higher speed models. The transition between both regions becomes steeper for lower injection velocities. We also note that deeper and more extended low-density regions are produced behind the shock wave as the injection velocity of the CME increases.

\begin{figure} [!h]
\centering
\includegraphics[width=0.9\textwidth]{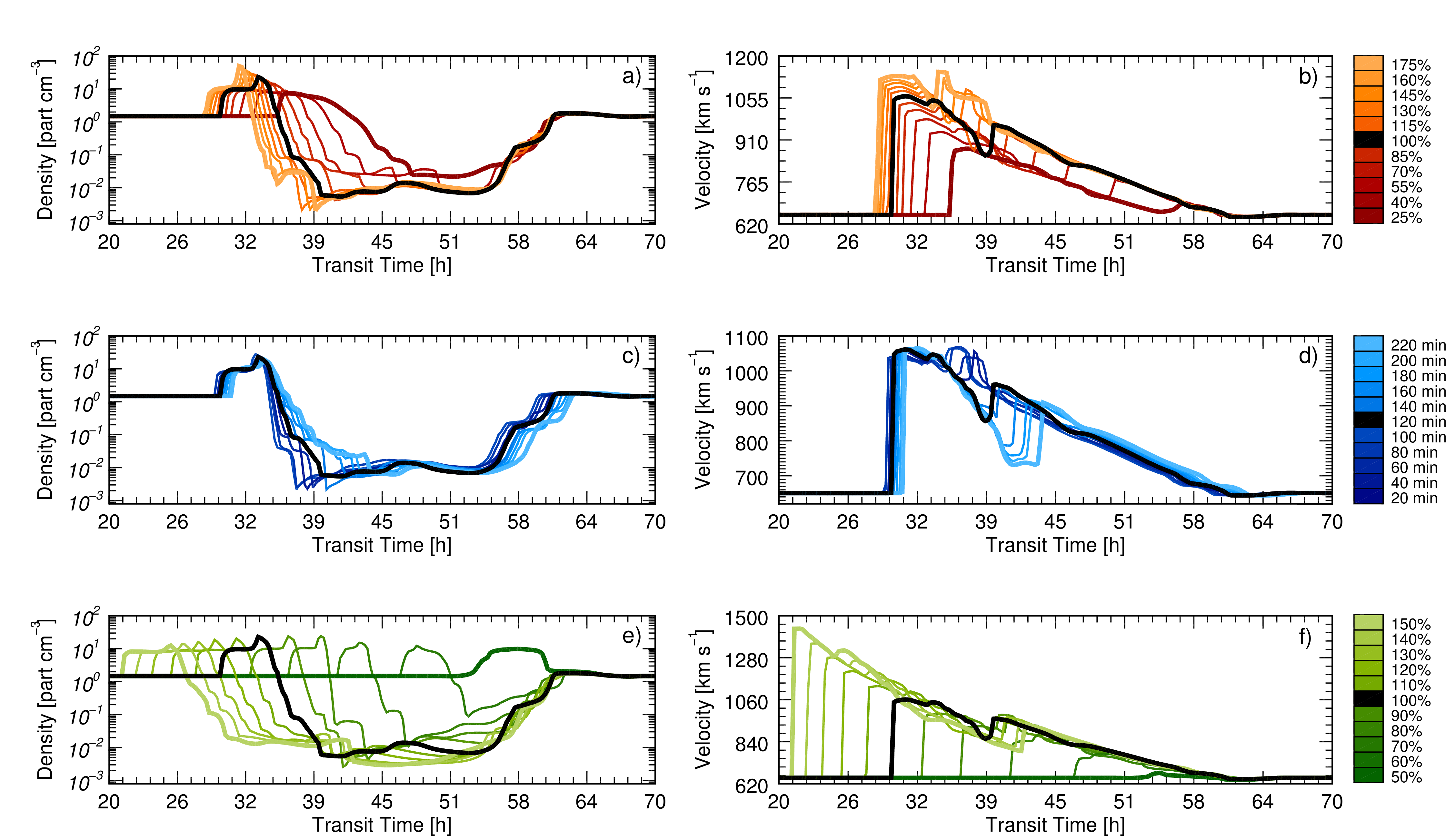} 
\caption{Predicted density and velocity profiles as function of time for numerical models with different input parameters (expelled mass $m$, duration $\Delta t$, and ejection velocity $v$) of the CME. The models are performed varying only one parameter, and the others are fixed. {\it Orange} profiles ({\it top panels}) are obtained from variations of $m$; {\it blue} profiles ({\it middle panels}) correspond to different values of $\Delta t$; and {\it green} profiles ({\it bottom panels}) are calculated changing $v$. The color bars at the right side show the variation of each parameter, and the {\it black lines} correspond to the predicted profiles of the reference model with $m=$ 1.1 $\times$ 10$^{16}$ g, $\Delta t=$ 2.0 h, and $v=$ 1330 km s$^{-1}$. See the text for further description.} 
\label{Fig:f2}
\end{figure}

As it can be seen in Figure 2,  it is remarkable the dependence of the model predictions with the remote sensing observations. In particular, variations of the injection velocity of the CME produce the major changes in the predicted density and velocity profiles. Nonetheless, we cannot arbitrarily change this parameter as the estimation of the measured velocity has the lowest observational error of $\pm$20 km s$^{-1}$ (Yashiro et al. 2004 $\&$ Xie et al. 2004). On the other hand, there are large discrepancies in the computation of the total mass of a CME with differences up to two orders of magnitude in the same event (as instance, see Stewart et al. 1974; Howard et al. 1985; Hundhausen et al. 1994; Colaninno et al. 2009; Vourlidas et al. 2010; Mishra et al. 2015). Therefore, it is possible to change this parameter in a wider range of values to improve the fit between the models and the observations at 1 AU. 

It is worth to mention that we did not perform simulations varying the angular width (or the solid angle $\Omega$) of the CME, due to the fact that the expelled mass $m = \dot m \, \Delta t \propto \Omega$, and therefore by varying $\Omega$ we have similar profiles to those obtained from changes of the total mass. Moreover, variations of the solid angle less than 0.17 rad produce effects in the predicted profiles that can be neglected.


\subsection{Models for different parameters of the solar wind}

\begin{figure} [!h]
\centering
\includegraphics[width=0.9\textwidth]{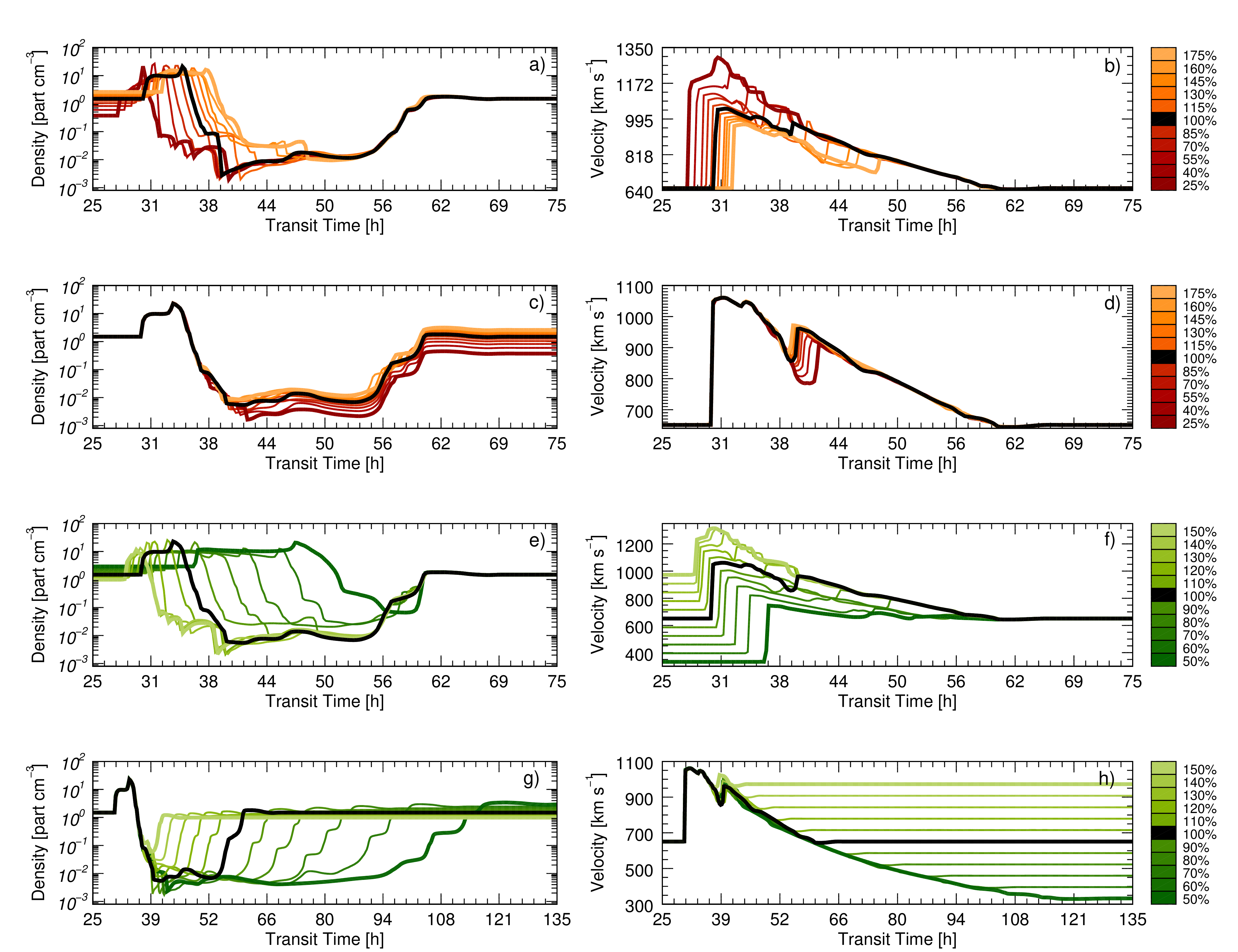} 
\caption{Predicted density and velocity profiles as function of time for numerical models with different solar wind conditions (mass-loss rate $\dot m$, and ejection velocity $v_{\mbox{\tiny SW}}$). These models are computed from variations of one parameter, and the other is fixed. {\it Orange} profiles correspond to variations of $\dot m$ of the pre-eruptive wind ({\it panels} a-b) and post-eruptive wind ({\it panels} c-d); and {\it green} profiles are obtained for different values of $v_{\mbox{\tiny SW}}$ of the pre-eruptive wind ({\it panels} e-f), and post-eruptive wind ({\it panels} g-h). The variations of each parameter are shown in the color bars at the right side of the figure. The reference models are depicted with {\it black lines}, which correspond to a solar wind with an ejection velocity $v_{\mbox{\tiny SW}}=$650 km s$^{-1}$ and $\dot m=$ 9.27 $\times$ 10$^{-15}$ M$_{\odot}$ y$^{-1}$. The detailed description of the figure is given in the text.} 
\label{Fig:f3}
\end{figure}

In order to investigate the dependence of the predicted density and velocity profiles at 1 AU with the initial conditions of the solar wind, we have performed numerical simulations varying the ejection velocity $v_{\mbox{\tiny SW}}$, and the mass-loss rate $\dot m$ of the wind. As a reference model, we have adopted  $v_{\mbox{\tiny SW}}=$ 650 km s$^{-1}$, and $\dot m=$ 9.27 $\times$ 10$^{-15}$ M$_{\odot}$ y$^{-1}$. Furthermore, in these simulations we have fixed the CME parameters, assuming an ejection speed $v=$1330 km s$^{-1}$, and an expelled mass $m=$ 1.1 $\times$ 10$^{16}$ g, during an interval of time $\Delta t=$2.0 h. The resulting profiles are presented in Figure \ref{Fig:f3}), in which the {\it black lines} depict the reference model.

First, we assume that the mass-loss rate of the pre-eruptive wind (Fig. \ref{Fig:f3}: {\it panels} a-b), and the corresponding value of the post-eruption wind (Fig. \ref{Fig:f3}: {\it panels} c-d), changes by $\pm$15\%, $\pm$30\%, $\pm$45\%, $\pm$60\%, and $\pm$75\%, with respect to the reference parameter $\dot m=$ 9.27 $\times$ 10$^{-15}$ M$_{\odot}$ y$^{-1}$. The predicted density and velocity profiles at 1 AU are presented. The models with the lowest value of the mass-loss rate ($\dot m=$ 2.31 $\times$ 10$^{-15}$ M$_\odot$ yr$^{-1}$) are depicted with {\it dark-red lines}, while the models with the highest value (1.62 $\times$ 10$^{-14}$ M$_\odot$ yr$^{-1}$) are shown with {\it light-orange lines}. We can see from {\it panel} (a) that longer transit times occur as the mass-loss rate of the pre-eruptive wind increases, as it is expected since the pre-eruptive wind slows the ejection. Besides, more extended compression regions and less-deep density drops are observed. In addition, it is observed in {\it panel} (b) that as lower is the mass-loss rate of the wind, stronger shocks arrive at Earth, as well as the detection of the shock wave and the rarefaction zone last for longer times. On the other hand, variations in the mass-loss rate of the post-eruption wind show that both the density and velocity profiles ({\it panels} c-d) are very similar in all models. Then, this parameter has no significant effect on the dynamics of the shock structure, that is, the arrival time and speed of the leading shock are not modified. As it is expected, only the rarefied zone behind the compression region is affected.

Also, we compute numerical models assuming that the ejection velocity of the pre-eruptive wind (Fig. \ref{Fig:f3}: {\it panels} e-f), or the corresponding speed of the post-eruption wind (Fig. \ref{Fig:f3}: {\it panels} g-h), changes by $\pm$10\%, $\pm$20\%, $\pm$30\%, $\pm$40\%, and $\pm$50\% with respect to the reference value $v_{\mbox{\tiny SW}}=$ 650 km s$^{-1}$. We present in these {\it panels} the density and velocity as function of time at 1 AU predicted by our simulations. The models with the highest speed correspond to $v_{\mbox{\tiny SW}}=$ 950 km s$^{-1}$ ({\it light-green lines}), while those with the slowest speed correspond to $v_{\mbox{\tiny SW}}=$ 325 km s$^{-1}$ ({\it dark-green lines}). We note from the density profiles that as the pre-eruptive wind velocity decreases, the transit time of the leading shock is longer, as well as the compression region is more extended and the rarefied zone is denser. Moreover, the velocity profiles show stronger arrival shocks in the lower velocity models. On the other hand, the resulting profiles from variations of the velocity of the post-eruption solar wind show that the compression region is not modified in these simulations, but obviously, the rarefaction zone is more extended in the lower-velocity models. Consequently, observations at 1 AU of the ICME structure, which consists of the shock front and the compression region, might not depend on the post-eruption solar wind conditions.
 
We note from our simulations that the density in the the rarefaction zone reaches very low values. This is a limitation of our model, the density may reach higher values in this region whether the magnetic field is included.


\subsection{Parametrization}

\begin{figure} [!p]
\centering
\includegraphics[width=0.8\textwidth]{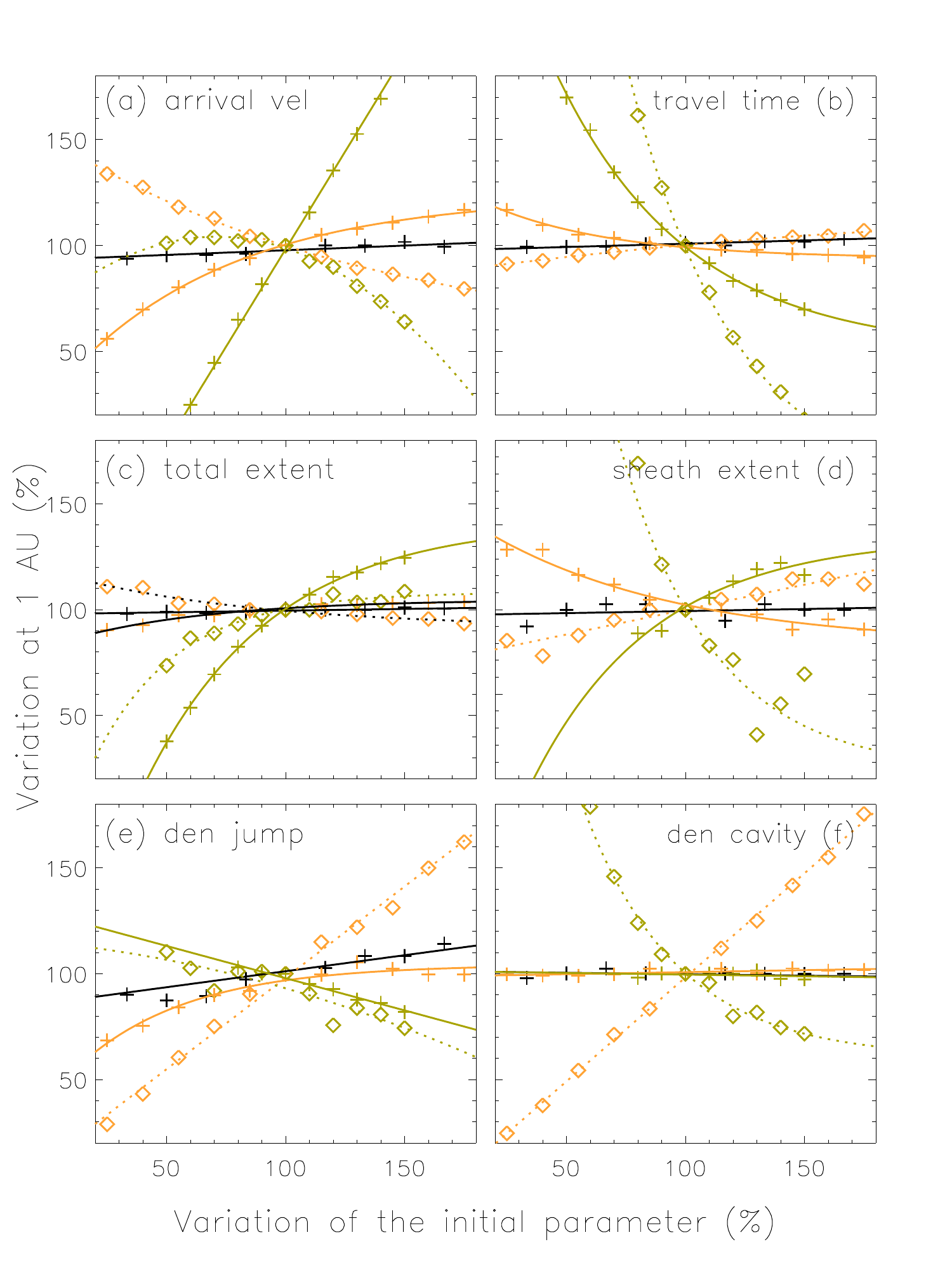} 
\caption{Numerical results of the ICME structure at 1 AU as function of the initial parameters. Percentages of variation of arrival velocity ({\it panel} a), travel time ({\it panel} b), total extension ({\it panel} c), compression region extension ({\it panel} d), density jump at the shock front ({\it panel} e), and density drop in the rarefied zone ({\it panel} f) are presented. Changes as function of the CME injection parameters are shown with plus symbols connected with {\it solid lines}, while diamonds connected with {\it dotted lines} correspond to variations as function of the solar wind conditions. In both cases, variations of the mass-loss rate and injection velocity are depicted in {\it orange} and {\it green}, respectively. Plus symbols connected with black {\it solid lines} show the results obtained by changing the duration of the CME.} 
\label{Fig:f4}
\end{figure}

Based on the previous exercise, we investigate possible correlations between the physical properties of the ICME at 1 AU and the injection parameters of the CME. Our results are summarized in Figure \ref{Fig:f4}, where we present (in percentage) the variations of the arrival velocity ({\it panel} a), the travel time ({\it panel} b), the total extension ({\it panel} c), the compression region extension ({\it panel} d), the density enhancement ({\it panel} e), and the density drop ({\it panel} f) of the ICME as function of the injection parameters of the CME. \footnote{We call total extension, the interval of time in which the parameters of SW have changed, which includes the compression region and the rarefaction zone. The interval of time starts with the arrival at 1 AU of the leading shock and ends when the SW conditions resume again. This extension can be seen clearly in the velocity profile, with the arrival of the shocked structure driven by the ICME, when the speed suddenly increases and then, returns back to the SW speed. Whereas the compression region extension is the interval of time delimited by the increase of  density. It starts with leading shock and ends when this parameter returns to the value of the ambient solar wind.}

In the figure, diamonds (connected by {\it dotted lines}) correspond to variations of the initial solar wind parameters, and plus symbols (connected by {\it solid lines}) display variations of the CME parameters. Besides, the predicted results obtained varying the mass-loss rate are depicted in {\it orange}, while those obtained from changes of the velocity are shown in {\it green} color. Black plus symbols connected by {\it solid lines} represent variations of the CME duration. 

Important features can be seen in the figure. We note first that the CME duration has no significant effect on the arrival parameters of the ICME, and consequently, the {\it black solid lines} are roughly flat in all {\it panels}. This happens because the total mass of the CME and speed remain fix for all the simulations in which the duration time was changed, therefore, the total momentum is the same for all cases.

 When possible, we have fitted analytical expressions to the relationship between CME and ICME variations in percentage. The variations of the arrival velocity depend on changes of the CME parameters (speed and mass-loss rate, {\it panel} a) as follows, $\Delta v_{\mbox{\tiny ARR}} = -85.92 +  1.84 ~\Delta v$, and $\Delta v_{\mbox {\tiny ARR}} = -95.96 ~\exp(-0.014 ~\Delta \dot{m}) + 123$, respectively, where the mass-loss rate of the CME is defined as $\dot{m} \equiv m/ \Delta t$. On the other hand, by varying the solar wind conditions, we found: $ \Delta v = 71.46 + 0.92\Delta v_{\mbox{\tiny SW}} - 0.01\Delta v_{\mbox{\tiny SW}}^2 $, and $ \Delta v = 98.85 \exp( -0.007 \Delta \dot{m}_{\mbox{\tiny SW}} ) +  53$. It is noteworthy that there is an asymptotical behavior of the arrival velocity with a limit of $\sim 23 \%$ when the CME mass increases. For the travel time ({\it panel} b), we found that it is more affected by variations of the injected velocity of both the solar wind, $ \Delta TT =  1087.68\, \exp(- 0.022 \,\Delta v_{\mbox{\tiny SW}} ) - 19$, and the CME, $ \Delta TT =  299.93\, \exp(- 0.0180\, \Delta v) + 50$).

The total extension of the ICME structure $t_{\mbox{\tiny EXT}}$ ({\it panel} c) is a function of $v$ and $v_{\mbox{\tiny SW}}$ as $\Delta t_{\mbox {\tiny EXT}} =-264.70\, \exp(-0.018 \Delta v) + 142$, and $\Delta t_{\mbox{\tiny EXT}} =  -138.61\, \exp( -0.028 \Delta v_{\mbox{\tiny SW}}) + 108$), with asymptotic values of 42$\%$ and 8$\%$, respectively. The compression region extension $s_{\mbox{\tiny EXT}}$ ({\it panel} d) is more influenced by $v_{\mbox{\tiny SW}}$ and $\dot{m}$ as: $\Delta s_{\mbox{\tiny EXT}} = 1205.85\, \exp(- 0.025 \Delta v_{\mbox{\tiny SW}} ) +  4$, and $ \Delta s_{\mbox{\tiny EXT}} =  82.24\, \exp(- 0.012 \Delta \dot{m}) + 79$. Furthermore, we found that the density jump $n_{\mbox{\tiny J}}$ ({\it panel} e), when the shock front arrives at 1 AU, depends on the variations of the solar wind conditions as: $\Delta n_{\mbox{\tiny J}} = 11.73 + 0.86\,\Delta \dot{m}_{\mbox{\tiny SW}}$, and $\Delta n_{\mbox{\tiny J}} = 114.742 - 0.111\,\Delta v_{\mbox{\tiny SW}} -0.001 \,\Delta v_{\mbox{\tiny SW}}^2$. For the CME parameters, a good fit between the injection speed and the density jump is given by $\Delta n_{\mbox{\tiny J}} = 128.32 - 0.30\,\Delta v$. It is interesting that the density jump is not affected at higher values of the expelled mass of CME. Finally, the density drop $n_D$ in the rarefied zone depends on the solar wind parameters as $\Delta n_{\mbox{\tiny D}} = -0.05 + 0.98\,\Delta \dot{m}_{\mbox{\tiny SW}}$, $\Delta n_{\mbox{\tiny D}} = 500.14\,\exp(0.025 \Delta v_{\mbox{\tiny SW}} ) + 60$, but no dependence with the injection parameters of the CME is obtained from the simulations.

The ICME morphology is very important in terms of space weather. For instance, the compression region extension, speed and density jumps may be used to predict sudden storm commencement of geomagnetic storms (Huttunen et al. 2005 $\&$ Despirak et al. 2009). Therefore, this exercise may help predict the effects of a CME at the Earth environment. Although, it is necessary to know the information of the magnetic field strength and polarity to predict the geomagnetic storm completely. 


\subsection{Comparison with observations}

We apply the parametric study presented above in order to find the best fit to the {\it in-situ} observations of the event of July 25, 2004 ({\it black solid line}). The ICME starting time was July 26, 2004 22:25 and the ending time was July 27, 2004 23:59 according to the Wind ICME Catalogue ($http://wind.nasa.gov/index\_WI\_ICME\_list.htm$), corresponding to the ICME duration of 25.56 h. 

\begin{figure} [!h]
\centering
\includegraphics[scale=0.68, angle=90]{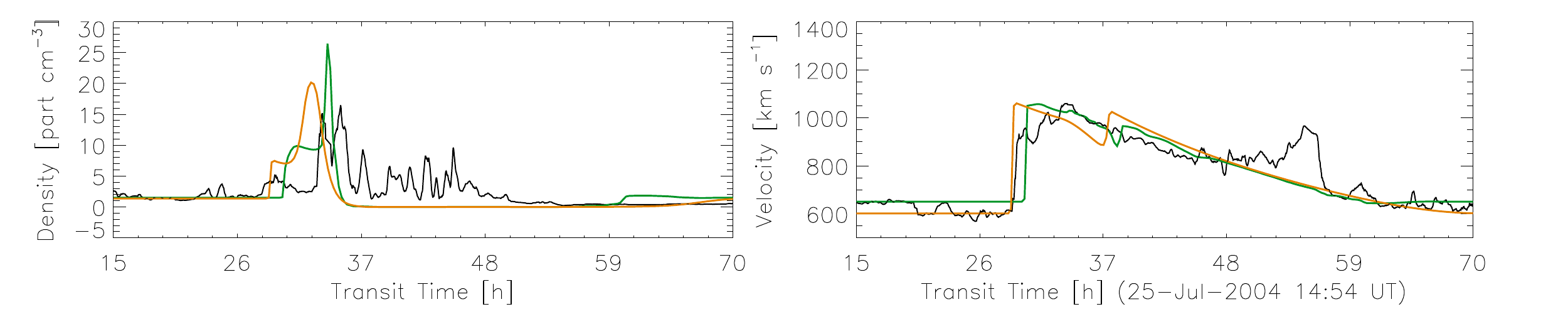} 
\caption{Profiles of density ({\it left panel}) and velocity ({\it right panel}) as function of time for two different models. The model MEDIUM is shown in {\it green} while in {\it orange} a model in which we reduced the solar wind speed $\sim +8 \%$ and increased the CME mass-loss rate $\sim +50 \%$. The {\it in situ} data obtained by WIND are shown in {\it black lines} with a smooth of 15 minutes which is our computational time.} 
\label{Fig:f5}
\end{figure}

In Figure \ref{Fig:f5}, we show the density and velocity profiles obtained by two different models: the model MEDIUM (see, Table 1) is presented with the {\it green line}, whereas our best fit to the observations is depicted with the {\it orange line}. We have adopted the initial solar wind parameters $v_{\mbox{\tiny SW}}=$ 595 km s$^{-1}$ and $\dot{m}_{\mbox{\tiny SW}}=$ 1.28 $\times$ 10$^{-13}$ M$_{\odot}$ yr$^{-1}$, obtained by reducing by $\sim 8\%$ and increasing by $\sim 50 \%$ the original values of the model MEDIUM, respectively. We note that our model predicts a duration of 27.5 h, that is, less than two hours of difference with the observed value. In terms of the duration of the compression region, our numerical model predicts a duration of 4.5 h, and then, we get a difference with observations of 1.72 h. In addition, we have a difference in the transit time $\Delta TT < $ 0.1 h. Hence, our numerical results are in agreement with observations at 1 AU. Nevertheless, it is worth to mention that our predicted density profile, as well as the one obtained from the model MEDIUM, shows a rarefaction zone that is not consistent with observations. 

This discrepancy between the predicted and observed profiles results from assuming the same conditions in both the pre-eruptive and post-eruption winds. However, it was shown in our parametric study, that the rarefaction zone is more extended as the velocity of the post-eruption wind decreases (see, Fig. 3). The duration of the complete structure enlarges or shortens depending on the conditions of the post-eruption solar wind. This is very important to consider when characterizing the duration and structure of the ICMEs.

\begin{figure} [!h]
\centering
\includegraphics[scale=0.68, angle=90]{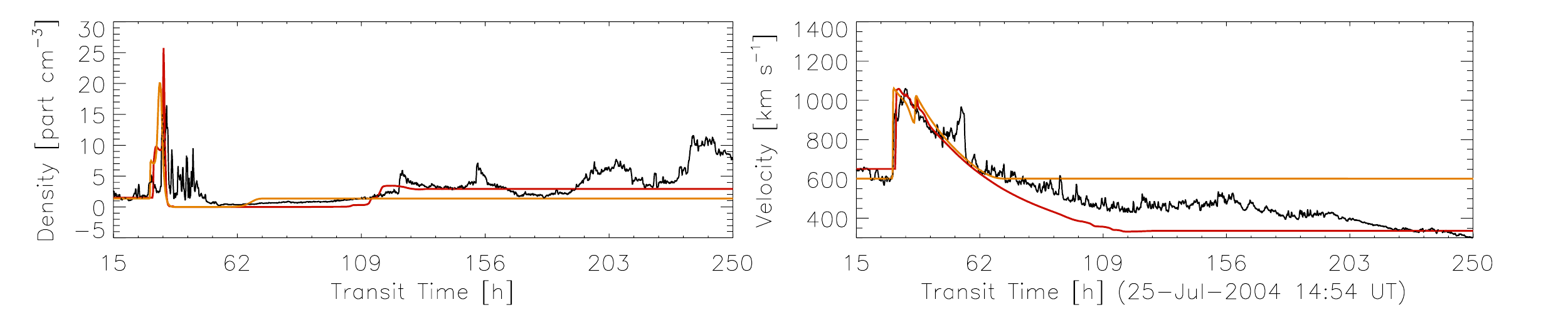} 
\caption{Profiles of density ({\it left panel}) and velocity ({\it right panel}) as function of time for two different models. Our best fit model is shown in {\it orange} while in {\it red} a model in which the post-eruption solar wind was slower down to $v_{\mbox{\tiny SW2}} =$325 km s$^{-1}$. It results in a better fit of the rarefaction zone. The {\it in situ} data obtained by WIND are shown in {\it black lines} with a smooth of 15 minutes.} 
\label{Fig:f6}
\end{figure}

In Figure \ref{Fig:f6}, we show in {\it orange} the best fit to observation while in red a new model in which the
speed of the post-eruption solar wind has been decreased to $v_{\mbox{\tiny SW2}}=$325 km s$^{-1}$.

\section{Numerical simulations of ICME-ICME interaction events}

We studied three different events (May 23, 2010; August 1, 2010; and November 9, 2012) on which two successive CMEs were launched in similar directions into the IPM. Using the model described in $\S$2, we investigate through numerical simulations the ICME-ICME interactions and their dynamical evolution.


\subsubsection{Event of May 23, 2010}
\label{may23}

LASCO (on board of {\it SoHO} spacecraft; Brueckner et al. 1995) detected the first eruption (CME$_1$) on May 23, 2010 at 18:06 UT, with a velocity $v_{1} = 400$ km s$^{-1}$, and a total mass $m_{1}= 1.5 \times 10^{16}$ g. The estimated duration of the eruption (from its brightness distribution; see Lara et al. 2004) was $\Delta t_1 = 2.85$ h. Using these parameters, the computed mass-loss rate during the CME$_1$, within a solid angle $\Omega_1/4\pi \simeq 0.07$, is $\dot{m_1}= 3.46 \times 10^{-13}$ M$_\odot$ yr$^{-1}$. The second eruption CME$_2$ was observed by LASCO on May 24, 2010 14:06 UT, therefore, the interval of time between the two eruptions was $\Delta t_{\mbox{\tiny SW}} = 17.15$ h. The velocity and the computed mass for this eruption are $v_{2} = 650$ km s$^{-1}$ and $m_2 = 1.0 \times 10^{16}$ g, respectively. We estimated a duration $\Delta t_2 = 3.71$ h, and consequently, a mass-loss rate $\dot{m_2} = 1.32 \times 10^{-13}$ M$_\odot$ yr$^{-1}$ within a solid angle $\Omega_2/4\pi \simeq 0.09$.  For the solar wind, we have assumed a velocity  $v_{\mbox{\tiny SW}} = 320$ km s$^{-1}$ (that corresponds to the measured {\it in situ} value by WIND spacecraft on May 28, 2010 01:00 UT), and a mass-loss rate $\dot m_{\mbox{\tiny SW}} = 2 \times 10^{-14}$ M$_\odot$ yr$^{-1}$ (Wood et al. 2002; Cranmer 2004).

\begin{figure} 
\centering
\includegraphics[scale=0.68, angle=90]{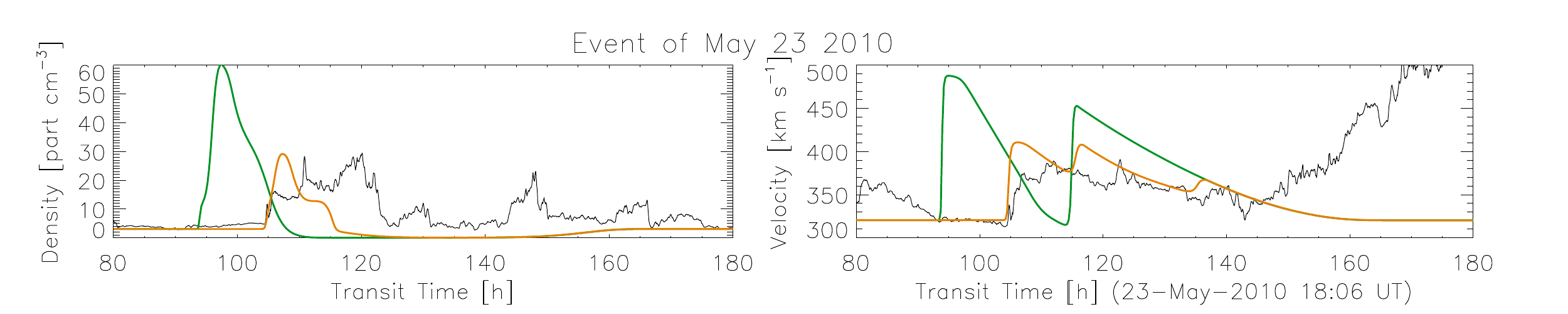} 
\caption{Predicted density ({\it left panel}) and velocity ({\it right panel}) profiles at 1 AU of the event of May 23, 2010. Observations {\it in situ} by WIND spacecraft are also shown ({\it black line}). The numerical simulation assuming the reported parameters of the two CMEs corresponds to the {\it green line}, whereas the numerical model depicted with the {\it orange line} is obtained by reducing 75$\%$ the expelled mass during the eruptions, and 20 km s$^{-1}$ the ejection velocities of the CMEs. We have smoothed observations of the event every 15 min, in order to be consistent with the numerical simulations. Further description of the figure is given in the text.}
\label{Fig:f7}
\end{figure}

In Figure \ref{Fig:f7}, we present in {\it green} the density ({\it left panel}) and the velocity ({\it right panel}) profiles at 1 AU of our numerical model. As it was shown in $\S$3.3, it is possible to get a better fit to observations varying some of the initial conditions. We have reduced by 75$\%$ the CMEs mass-loss rates ($\dot{m_1}= 8.65 \times 10^{-14}$, and $\dot{m_2}= 3.3 \times 10^{-14}$), and also, by 20 km s$^{-1}$ the CME speeds ($v_{1} = 380$ km s$^{-1}$, and $v_{2} = 630$ km s$^{-1}$). Assuming these changes, we obtain the profiles depicted with {\it orange lines} in the figure. From the comparison with observations, we obtain a difference between the observed and predicted arrival velocities $<$ 10 km s$^{-1}$, while the arrival density differs $<$ 10 cm$^{-3}$ with respect to the observed value. In addition, we obtain a difference $<$ 0.1 h for the arrival time.

\begin{figure} 
\centering
\includegraphics[width=0.9\textwidth]{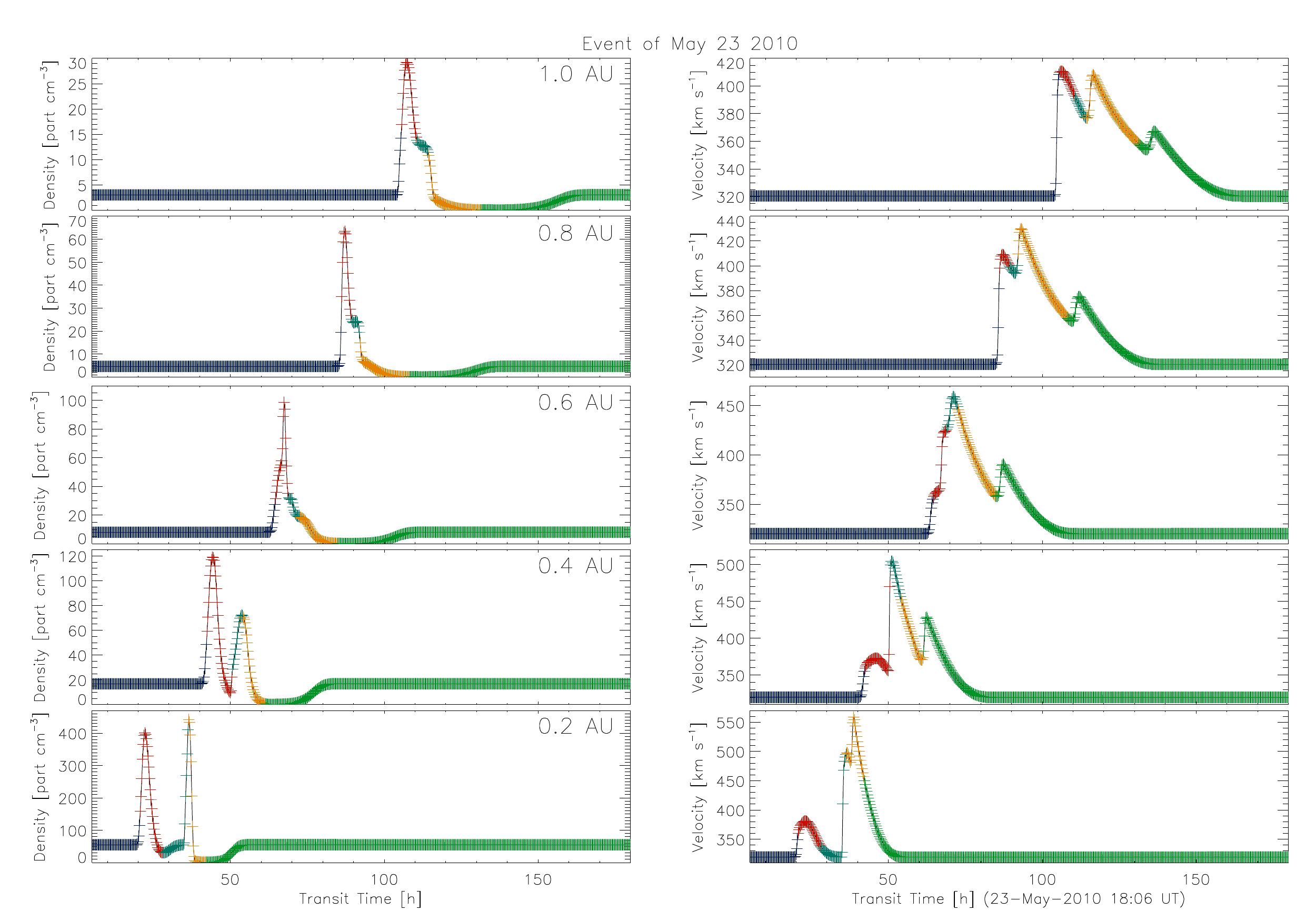} 
\caption{Profiles of density ({\it left panels}) and velocity ({\it right panels}) as function of time for different heliospheric distances (from {\it bottom} to {\it top}: 0.2 AU, 0.4 AU, 0.6 AU, 0.8 AU and 1.0 AU) for the event of May 23, 2010. The interaction took place at a distance of $R\simeq$ 0.8 AU (which corresponds to a time of $\simeq$ 77.29 h). According to our numerical model, the complex structure formed after the collision arrived to the Earth $\simeq$97.5 h after the first CME was launched. The model predicts an arrival ICME speed of 417 km s$^{-1}$). We tagged each plasma with crosses colored: the solar wind before CME$_1$ in {\it blue}, the CME$:_1$ in red, the solar wind between the CMEs in aqua, the CME$_2$ in yellow and the solar wind after the CME$_2$ in {\it green}.}
\label{Fig:f8}
\end{figure}

In Figure \ref{Fig:f8},  we present the density ({\it left panels}) and velocity ({\it right panels}) profiles, at heliospheric distances $R =$ 0.2, 0.4, 0.6, 0.8 and 1.0 AU (from {\it bottom} to {\it top}) predicted by our model for the event of May 23, 2010. In the figure, we have tagged the different plasmas with colored plus symbols as follows: the pre-eruptive wind (in {\it blue}), the first CME (in {\it red}), the solar wind between the eruptions (in {\it aqua}), the second CME (in {\it yellow}), and the post-eruption wind (in {\it green}). 

It is noteworthy that the third shock predicted in the velocity profile may be due to the lack of the magnetic field which inhibits the compression process. In the observed profiles, the presence of the magnetic field vanish any possible shock structure inside the rarefaction zone.

We predict the ICME-ICME collision at a time $t \simeq$ 77.29 h after the first CME is launched, and at a distance $R\simeq$ 0.8 AU from the Sun. Beyond this distance, a single merged region is formed containing the material expelled during both eruptions. Afterwards, this region decelerates with time, reaching the Earth at a time $t_{\mbox{\tiny ARR}} \simeq$ 97.5 h, with an arrival speed $v_{\mbox {\tiny ARR}} =$ 417 km s$^{-1}$. These numerical predictions are consistent with the observed arrival time of $\simeq$101 h, and the measured speed {\it in situ} of $\simeq$ 380 km s$^{-1}$ (see, for instance, Lugaz et al. 2012).

Moreover, our numerical simulations are in good agreement with analytic results reported by Niembro et al. (2015) for the May 23, 2010 event. These authors predicted that the ICME-ICME interaction occurred at a time $\simeq$ 75.08 h, and at a distance $\simeq$ 0.69 AU from the Sun. Additionally, the predicted arrival time and velocity of the merged region are $\simeq$ 105.19 h, and $\simeq$ 427 km s$^{-1}$, respectively.

We note that in Lugaz et al. (2012), it is argued that the CME$_2$ did not reach WIND spacecraft because of its deflection from the interaction with the CME$_1$. They stated that the density and velocity profiles observed are only due to the arrival of CME$_1$. Nevertheless, we assumed that both CMEs were ejected in the same direction and the deflection did not take place. Our simulation reproduces well the observations and shows that there are features corresponding to both CMEs. Consequently, that the structure observed by the spacecraft at 1 AU may include material from both CMEs. A more detailed analysis of this event is required.

\begin{figure} 
\centering
\includegraphics[scale=0.68, angle=90]{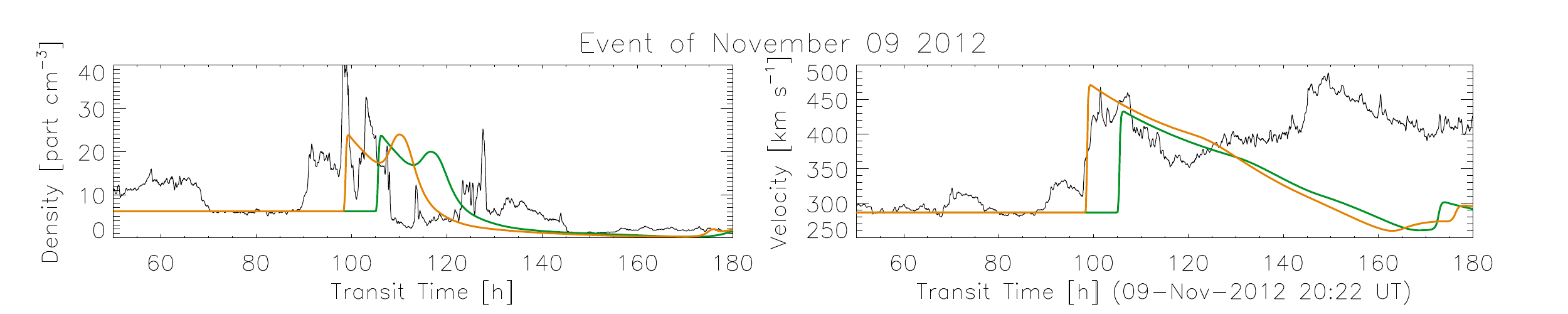} 
\caption{Same as Fig. 7, but for the event of November 9, 2012. The mass-loss rates of the pre-eruptive wind has been increased 20$\%$, the expelled mass of the CMEs 50$\%$, and their speeds 20 km s$^{-1}$. The discussion of the figure
is given in the text.}
\label{Fig:f9}
\end{figure}


\subsubsection{Event of November 9, 2012}

On November 9, 2012 (see Mishra et al. 2015 for more details of this event), two successive earth-directed CMEs were launched and detected by \emph{STEREO/SECHI} (Howard et al. 2008). The CME$_1$ was expelled from the Sun on November 9, 2012 at 17:39 UT, while the CME$_2$ was observed on November 10, 2012 at 06:39 UT. For CME$_1$, we assume that a mass of $m_{1}= 4.66 \times 10^{15}$ g was expelled during $\Delta t_1\simeq$1.55 h, with a mean velocity of $v_{1} = 500 $ km s$^{-1}$.  The estimated mass-loss rate of the eruption is $\dot{m_1}= 9.04 \times 10^{-14}$ M$_\odot$ yr$^{-1}$, within a solid angle $\Omega_1/4\pi = 0.14$. During the CME$_2$, a mass of  $m_2 = 2.27 \times 10^{15}$ g was expelled with a mean velocity of $v_{2} = 1100$ km s$^{-1}$. This second eruption last $\Delta t_2\simeq $2.5 h, obtaining a mass-loss rate of $\dot{m_2} = 4.42 \times 10^{-14}$ M$_\odot$ yr$^{-1}$ within a solid angle $\Omega_2/4\pi = 0.09$. For the solar wind, we have adopted an ejection velocity $v_{\mbox{\tiny SW}} = 300$ km s$^{-1}$ and a mass-loss rate $\dot m_{\mbox{\tiny SW}}= 1.5 \times 10^{-14}$ M$_\odot$ yr$^{-1}$.

In Figure \ref{Fig:f9}, we present the predicted profiles of density (\textit{left panel}) and velocity (\textit{right panel}) as function of time, assuming the above values ({\it green lines}). Also a comparison with the observed profiles ({\it black lines}) is shown. The observations are smoothed by intervals of time of 15 min, which corresponds to our numerical outputs. It can be seen from the figure that the merged region reaches the Earth at a time 105 h (a difference with the observed arrival time of more than 8 h), with a speed of 430 km s$^{-1}$, which differs only 15 km s$^{-1}$ from the {\it in situ} data, and a density of 23 part cm$^{-3}$, that differs more than 10 part cm$^{-3}$.

Our best fit to observations ({\it orange line} is obtained assuming a mass-loss rate of the solar wind $\dot m_{\mbox{\tiny SW}}= 1.8 \times 10^{-14}$ M$_\odot$ yr$^{-1}$, which is 20$\%$ higher than the original value. The mass-loss rates of the CME$_1$, and CME$_2$ are increased 50$\%$, thus $\dot m_{1}= 1.356 \times 10^{-13}$ M$_\odot$ yr$^{-1}$, and $\dot m_{2}= 6.63 \times 10^{-13}$ M$_\odot$ yr$^{-1}$, respectively. This model differs 25 km s$^{-1}$, and $<$ 10 part cm$^{-3}$ with respect to the arrival velocity and density of the merged region. The arrival time is exactly the same as the observations. 
 
\begin{figure} 
\centering
\includegraphics[width=0.9\textwidth]{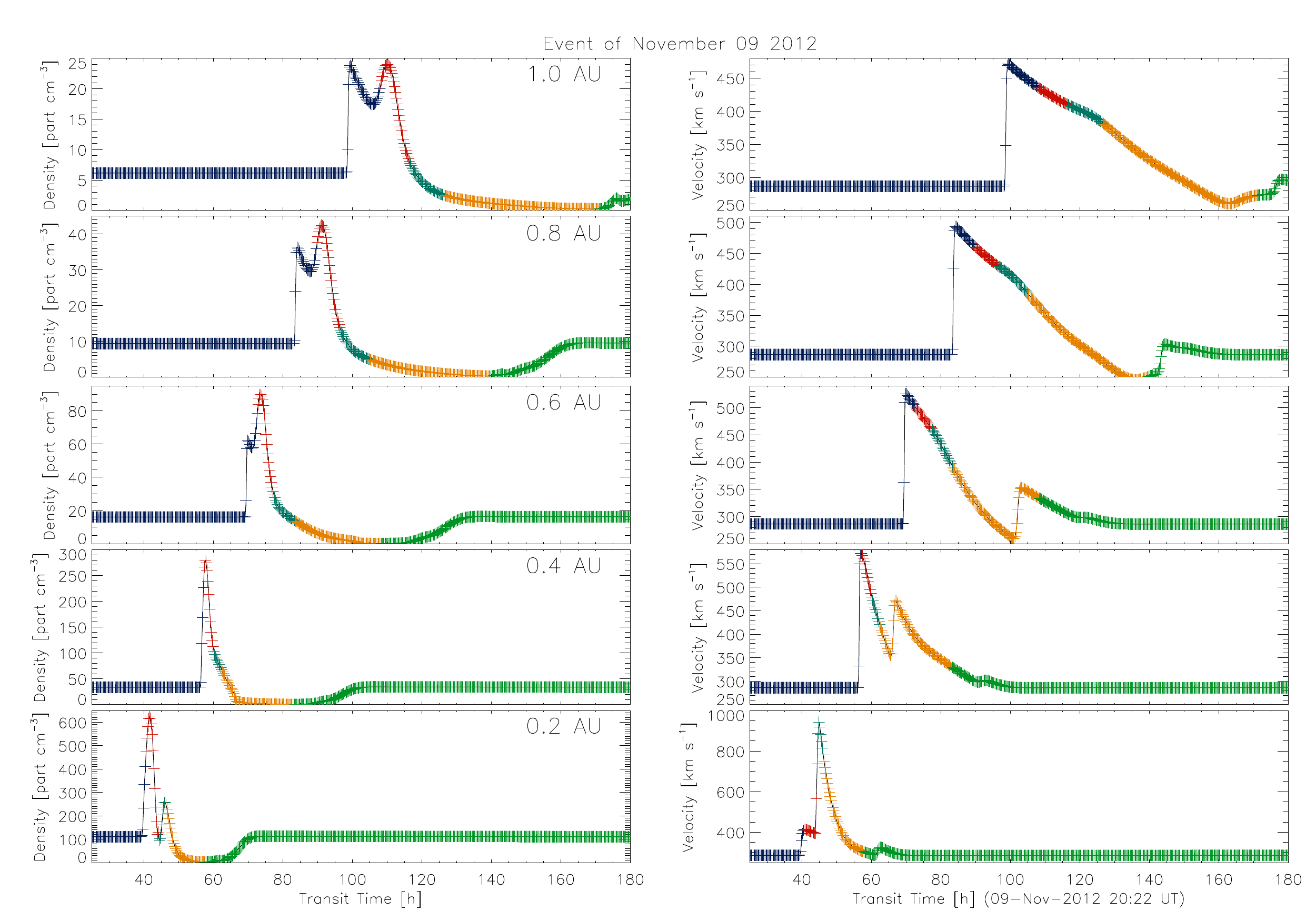} 
\caption{Same as Fig. 8, but for the event of November 9, 2012. In this event, the interaction is predicted
 at a distance of $R\simeq$ 0.18 AU, which corresponds to a time of $t \simeq$ 35 h. The predicted arrival time and velocity (at 1 AU) of the merged structure is 75 h after the first eruption, and 470 km s$^{-1}$, respectively.}
\label{Fig:f10}
\end{figure}

In Figure \ref{Fig:f10} we show the predicted time-sequence of density (left) and velocity (right) profiles for the event of November 9, 2012. The simulation predicts that the ICME-ICME interaction occurs at a time $\simeq$ 35 h after the first eruption, at a heliospheric distance of $\simeq$ 0.18 AU, and then, a merged region is formed. Therefore, these predictions differ from the observations by 2 h, and 0.2 AU, respectively. 

It is interesting to note that the evolution of the compression region after the interaction is variable. This is due to the fact that there is no interchange of material between both interacting ICMEs. In this case, the masses of the CMEs are similar in which changes in the two peaks of the density profiles are observed.

This event is reported by Mishra et al. (2015), in which three different components are identified within the arrival merged region at 1 AU: a shock front followed by a compression region, the CME$_1$, an interaction region, and the CME$_2$. We performed our simulations taking into account only the CMEs and not the interaction region. Our simulation shows that the CME$_2$ (v$_2=$ 1100 km s$^{-1}$) overtook the CME$_1$ (v$_1=$ 500km s$^{-1}$) at 0.18 AU (Figure 10), making impossible the existence of the Interaction Region at 1 AU. This event is a good example of how difficult is the identification of the distinct components of complex regions.

The numerical predictions are consistent with observational values, as well as with analytic predictions reported by Niembro et al. (2015).  The observations suggest that the collision occurred 19 - 36 h after the CME$_1$ was launched, at a distance of 0.16 - 0.46 AU. The observed arrival time of the merged structure was $\simeq$ 96 h, with a velocity $\simeq$ 450 km s$^{-1}$. On the other hand, Niembro et al. (2015) predict that the CMEs collide at $\simeq$ 34.97 h, when they are located at a distance $\simeq$ 0.32 AU. In their models, the merged region arrives to the Earth $\simeq$ 99.23 h after the first eruption, with a velocity of $\simeq$ 423 km s$^{-1}$.


\subsubsection{Event of August 1, 2010}

In the August 1, 2010 event, three successive CMEs were detected (e.g. Temmer et al. 2012; Harrison et al. 2012 and Liu et al. 2012). Nevertheless, this event has been frequently studied as the interaction of two consecutive CMEs. Here, we just consider the interaction between the last two ejections (here in after CME$_1$ and CME$_2$). The CME$_1$ was launched on August 1 2010 at 02:55 UT, with a speed of $v_{1} = 732$ km s$^{-1}$ during an interval of time of $\Delta t_1\simeq$1.1 h and a $m_{1}= 8.0 \times 10^{15}$ g, that result in a mass-loss rate $\dot{m_1}= 2.74 \times 10^{-13}$ M$_\odot$ yr$^{-1}$, within a solid angle $\Omega_1/4\pi = 0.12$.  The CME$_2$ was launched from the Sun on August 1 at 7:45 UT,  with the outflow parameters, $v_{2} = 1138$ km s$^{-1}$, and $m_2 = 3.0 \times 10^{16}$ g, within a solid angle $\Omega_2/4\pi = 0.18$.  This eruption last $\Delta t_2\simeq$ 1 h and, and thus, $\dot{m_2} = 7.39 \times 10^{-13}$ M$_\odot$ yr$^{-1}$. For the solar wind,  we assume an ejection velocity $v_{\mbox{\tiny SW}} = 410$ km s$^{-1}$, and a mass-loss rate $\dot m_{\mbox{\tiny SW}}= 2 \times 10^{-14}$ M$_\odot$ yr$^{-1}$  (see, for instance, Wood et al. 2002; Cranmer 2004).

\begin{figure} 
\centering
\includegraphics[scale=0.68, angle=90]{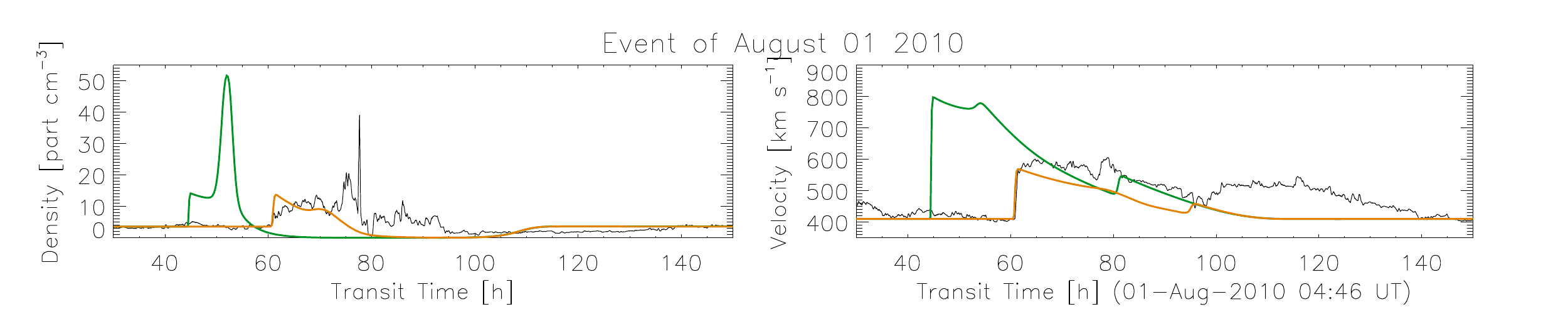} 
\caption{Same as Fig. 7, but for the event of August 1, 2010. The best fit is obtained reducing less than 35$\%$ the mass-loss rate of the CMEs, and less than 5$\%$ their initial speeds. The description of the figure is given in the text.}
\label{Fig:f11}
\end{figure}

Figure \ref{Fig:f11} shows the predicted profiles of density (\textit{left panel}) and velocity (\textit{right panel}).  as function of time (\textit{green lines}). Observations {\it in situ} by WIND spacecraft are also presented ({\it black line}), which are smoothed by intervals 15 minutes, as in the previous events. It can be seen in the figure that the numerical results are not consistent with observational data, since the arrival time and velocity differs more than 20 h and 200 km s$^{-1}$, respectively.

Our best fit model (\textit{orange line} is obtained with the input parameters of the CMEs, $v_{1} = 700$ km s$^{-1}$, $v_{2} = 1100$ km s$^{-1}$, $\dot{m_1} = 3.014 \times 10^{-13}$ M$_\odot$ yr$^{-1}$, $\dot{m_2} = 9.9765 \times 10^{-13}$ M$_\odot$ yr$^{-1}$. In this model, the arrival velocity differs $<$ 5 km s$^{-1}$, whereas the density difference is $\simeq$ 3 part cm$^{-3}$ with respect to the {\it in situ} data. 

\begin{figure} 
\centering
\includegraphics[scale=0.65]{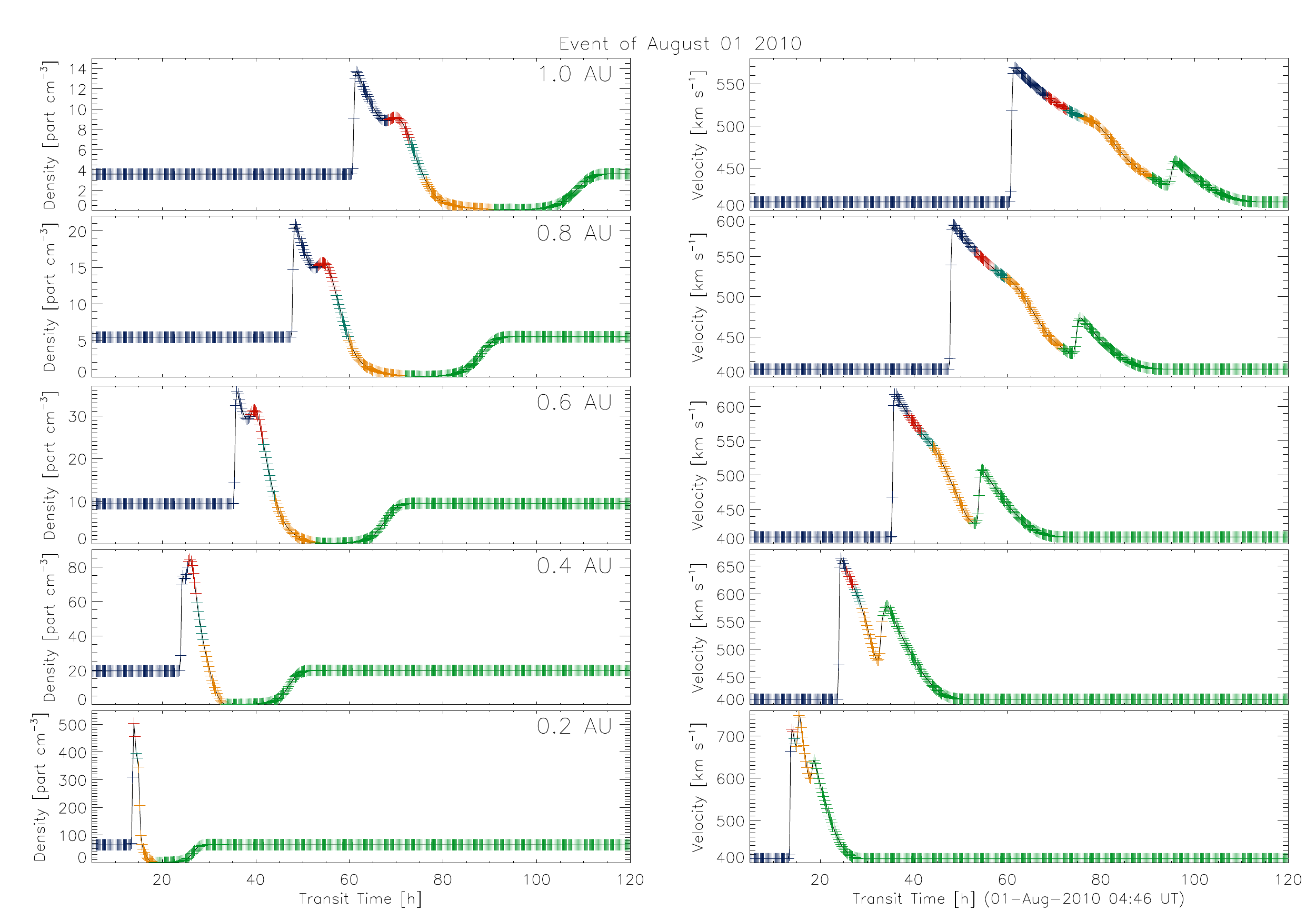} 
\caption{Same as Fig. 8, but for the event of August 01, 2010. In this event, the interaction is predicted at a distance of $R\simeq$ 0.19 AU at a time $\simeq$ 15 h after the first eruption. The merged region arrives to the Earth with a speed of 595 km s$^{-1}$, at an evolution time $\simeq$ 60.3 h}
\label{Fig:f12}
\end{figure}

In Figure \ref{Fig:f12}, we present a distance-sequence of density for the August 1, 2010 event. The density stratification at heliospheric distances $R =$ 0.2, 0.4, 0.6, 0.8 and 1.0 AU are shown. The numerical simulation predicts that ICME-ICME interaction occurs at a distance $\simeq$ 0.19 AU, which corresponds to an evolution time $\simeq$ 15 h after the first eruption. The merged region reaches the Earth with a speed of $\simeq$ 595 km s$^{-1}$, 60.3 h after the CME$_1$ is launched. We note that our numerical results are consistent with the observational data which suggest that the collision occurred at a time $\simeq$ 12.91 h after the first eruption, at a heliospheric distance $\simeq$ 0.16 AU. Besides, the observed arrival time of the merged region was $\simeq$ 60.3 h, with a velocity $\simeq$ 600 km s$^{-1}$. In addition, Niembro et al. (2015) obtained similar predictions for this event. They showed that the CMEs collide at $\simeq$ 13 h, when they are located at a distance $\simeq$ 0.2 AU. Also, these authors computed an arrival time of the merged region $\simeq$ 52.3 h, with a velocity of $\simeq$ 726 km s$^{-1}$.

\section{Conclusions}

In order to study the dynamics, evolution and time profile at 1 AU of CMEs traveling into the interplanetary medium, as a function of the initial conditions of both: the ambient medium and the CME,  we carried out a parametric study, using as reference the (single) CME event observed on July 25, 2004. By running numerical simulations varying the initial parameters, we found that the  CME time duration has the lowest influence on the ICME time profile morphology, while the parameter with major influence on this profile  is the CME velocity, followed by the CME mass-loss rate. This means that if we want to fit an observed ICME profile, we may vary the speed and/or the mass loss rate. Although, taking into account the observational uncertainties, we have $\sim 20\%$ of variation range over the speed, but we may vary the CME mass loss rate over a large range of values, because this is the parameter with the highest observational uncertainty, then, varying these parameters, we were able to reproduce the arrival time and the most important features of the speed and density profiles of our reference event.

In general, the compression region morphology depends on the pre-eruptive ambient solar wind and the CME parameters while the rarefaction zone structure depends on the CME parameters and the post-eruptive ambient solar wind conditions. With this in mind, we performed a simulation assuming a post-eruptive solar wind speed slower than the pre-eruptive one, and we were able to accurately predict the duration of the rarefaction zone. It is clear that the rarefaction is due not only to the presence of the magnetic cloud, but the conditions of the post-eruptive solar wind are very important for its duration. Therefore, this effect should be considered when characterizing the duration of the ICMEs.

Commonly, the studies of the ICME transport focus in the accurate prediction of the travel time and arriving velocity. Therefore, this parametric exercise is of utmost importance as it gives the relation of the speed and density profiles with the injection parameters, as well as clues of how the predictions of most of the hydrodynamic models can be improved by changing the CME mass-loss rate and/or the post-eruptive ambient solar wind conditions. 

We use the results of our parametric study to simulate the CME interaction events detected and tracked as they rushed outwards into space on May 23, 2010; August 1, 2010; and November 9, 2012. The YGUAZ\'U-A code used in our simulations, is able to tag the different flows involved in the interaction, i. e., the pre- and post-eruption ambient solar winds and both interacting CMEs. With this, we were able to identify the individual components (ICMEs) of the complex merged regions observed at 1 AU. This is an excellent tool to  analyze complex events and helps in its identification at 1 AU. 

In summary, following our parametrization method, we found that the best  predictions of the arrival time (matching the exact time of the observations) were obtained, in general, by sub-estimating the reported CME mass-loss rates. We found that our models can explain accurately the most important features of the speed and density profiles, and are able to distinguish, inside the merged region, the contribution of each of the interacting ICMEs. This study shows that the main characteristics of the ICME interaction can be obtained by studying the system hydrodynamically. 


\thanks{We would like to thank Alejandro Raga and Alejandro Esquivel for constructive discussions and all his helpful comments. This work was partially supported by CONACyT grants 344896, and 179588; and UNAM/PAPIIT grants IN111716-3, IN112014, IN112116 and IG100516.}
 


\begin{appendix} 
\section{The formation and evolution of a two-shock structure}

The interaction between two supersonic flows with increasing velocities must produce two shocks, since there is no way one shock can generate jump changes that can match the two flows (e.g. Pikelner 1968; Dyson $\&$ de Viers 1972). These authors pointed out that the injection of highly supersonic gas into the surroundings results in the formation of a shell between two shocks, with an outward shock that sweeps up the ambient medium and an inward shock that decelerates the high-velocity outflow. In principle, a contact discontinuity separates the shocked material of the surroundings and the shocked injected gas. This scenario has been studied for the impact of a stellar wind on the surrounding gas (eg. Dyson and Williams 1997), and also, on the origin of Planetary Nebulae, where a stellar wind from an evolved low-mass star interacts with the stellar wind of its previous phase of red giant (see, for instance, Kwok, Purton $\&$ Fitzgerald 1978). The same physical process occurs in the case of a Coronal Mass Ejection (CME) that interacts with the ambient solar wind. The onset of the CME is assumed to occur suddenly, during a finite interval of time, which is ejected with a higher speed than the standard solar wind. Consequently, the interaction between the CME and the ambient solar wind must produce a double-shock structure (see Raga et al. 1990; Cant\'o et al. 2005), which dynamical evolution can be studied using the parametric method developed by Cant\'o, Raga $\&$ D'Alessio (2000). For this time-dependent variability of the outflow phases (the CME and the solar wind), the shocked layer suffers two distinct stages on its dynamical evolution (see, Niembro et al. 2015, $\S$ 2.1). Initially, it propagates outwards with constant velocity, which is an intermediate value between the fast CME and the slower solar wind. Once all the mass expelled during the CME has entered the shocked shell, it is decelerated as it propagates into the slower solar wind. In this second final stage, the shell velocity steadily decreases asymptotically approaching the velocity of the ambient solar wind.

\end{appendix}



\begin{thebibliography}{}
\bibitem[Borgazzi et al.(2009)]{2009A&A...498..885B} Borgazzi, A., Lara, A., Echer, E., \& Alves, M.~V.\ 2009, A\&A, 498, 885 

\bibitem[Brueckner et al.(1995)]{1995SoPh..162..357B} Brueckner, G.~E., Howard, R.~A., Koomen, M.~J., et al.\ 1995, Sol. Phys., 162, 357 

\bibitem[Burlaga et al.(2002)]{2002JGRA..107.1266B} Burlaga, L.~F., Plunkett, S.~P., \& St.~Cyr, O.~C.\ 2002, Journal of Geophysical Research (Space Physics), 107, 1266 

\bibitem[Burlaga et al.(2003)]{2003JGRA..108.1425B} Burlaga, L., Berdichevsky, D., Gopalswamy, N., Lepping, R., \& Zurbuchen, T.\ 2003, Journal of Geophysical Research (Space Physics), 108, 1425 

\bibitem[Cant{\'o} et al.(2000)]{2000MNRAS.313..656C} Cant{\'o}, J., Raga, A.~C., \& D'Alessio, P.\ 2000, MNRAS, 313, 656 

\bibitem[Cant{\'o} et al.(2005)]{2005MNRAS.357..572C} Cant{\'o}, J., Gonz{\'a}lez, R.~F., Raga, A.~C., et al.\ 2005, MNRAS, 357, 572 

\bibitem[Colaninno \& Vourlidas(2009)]{2009ApJ...698..852C} Colaninno, R.~C., \& Vourlidas, A.\ 2009, ApJ, 698, 852 

\bibitem[Cranmer(2004)]{2004AmJPh..72.1397C} Cranmer, S.~R.\ 2004, American Journal of Physics, 72, 1397 

\bibitem[Despirak et al.(2009)]{2009AnGeo..27.1951D} Despirak, I.~V., Lubchich, A.~A., Yahnin, A.~G., Kozelov, B.~V., \& Biernat, H.~K.\ 2009, Annales Geophysicae, 27, 1951 

\bibitem[Dyson \& de Vries(1972)]{1972A&A....20..223D} Dyson, J.~E., \& de Vries, J.\ 1972, A\&A, 20, 223 

\bibitem[Dyson \& Williams(1997)]{1997pism.book.....D} Dyson, J.~E., \& Williams, D.~A.\ 1997, The physics of the interstellar medium.~ Edition: 2nd ed.~Publisher: Bristol: Institute of Physics Publishing, 1997.~Edited by J.~E.~Dyson and D.~A.~Williams.~Series: The graduate series in astronomy.~ISBN: 0750303069

\bibitem[Falkenberg et al.(2010)]{2010SpWea...8.6004F} Falkenberg, T.~V., Vr{\v s}nak, B., Taktakishvili, A., et al.\ 2010, Space Weather, 8, S06004 

\bibitem[Gonz{\'a}lez \& Cant{\'o}(2002)]{2002ApJ...580..459G} Gonz{\'a}lez, R.~F., \& Cant{\'o}, J.\ 2002, ApJ, 580, 459 

\bibitem[Gonz{\'a}lez et al.(2004)]{2004ApJ...600L..59G} Gonz{\'a}lez, R.~F., de Gouveia Dal Pino, E.~M., Raga, A.~C., \& Velazquez, P.~F.\ 2004, ApJl, 600, L59 

\bibitem[Gonz{\'a}lez et al.(2004)]{2004ApJ...616..976G} Gonz{\'a}lez, R.~F., de Gouveia Dal Pino, E.~M., Raga, A.~C., \& Vel{\'a}zquez, P.~F.\ 2004, ApJ, 616, 976 

\bibitem[Gonz{\'a}lez et al.(2010)]{2010MNRAS.402.1141G} Gonz{\'a}lez, R.~F., Villa, A.~M., G{\'o}mez, G.~C., et al.\ 2010, MNRAS, 402, 1141 

\bibitem[Gopalswamy et al.(2001)]{2001AGUFMSH12A0735G} Gopalswamy, N., Yashiro, S., von Rosenvinge, T.~T., \& Leske, R.\ 2001, AGU Fall Meeting Abstracts,  

\bibitem[Gopalswamy(2016)]{2016GSL.....3....8G} Gopalswamy, N.\ 2016, Geoscience Letters, 3, 8 

\bibitem[Harrison et al.(2012)]{2012ApJ...750...45H} Harrison, R.~A., Davies, J.~A., M{\"o}stl, C., et al.\ 2012, ApJ, 750, 45 

\bibitem[Howard et al.(1985)]{1985JGR....90.8173H} Howard, R.~A., Sheeley, N.~R., Jr., Michels, D.~J., \& Koomen, M.~J.\ 1985, JGR, 90, 8173 

\bibitem[Howard et al.(2008)]{2008SSRv..136...67H} Howard, R.~A., Moses, J.~D., Vourlidas, A., et al.\ 2008, SSR, 136, 67 

\bibitem[Hundhausen et al.(1994)]{1994ESASP.373..409H} Hundhausen, A.~J., Stanger, A.~L., \& Serbicki, S.~A.\ 1994, Solar Dynamic Phenomena and Solar Wind Consequences, the Third SOHO Workshop, 373, 409 

\bibitem[Hundhausen(1999)]{1999mfs..conf..143H} Hundhausen, A.\ 1999, The many faces of the sun: a summary of the results from NASA's Solar Maximum Mission., 143 

\bibitem[Huttunen et al.(2005)]{2005AnGeo..23..625H} Huttunen, K.~E.~J., Schwenn, R., Bothmer, V., \& Koskinen, H.~E.~J.\ 2005, Annales Geophysicae, 23, 625 

\bibitem[Kay \& Opher(2015)]{2015ApJ...811L..36K} Kay, C., \& Opher, M.\ 2015, ApJl, 811, L36 

\bibitem[Kwok et al.(1978)]{1978ApJ...219L.125K} Kwok, S., Purton, C.~R., \& Fitzgerald, P.~M.\ 1978, ApJl, 219, L125 

\bibitem[Lara et al.(2004)]{Lara2004} Lara, A., Gonzalez-Esparza, J.~A., \& Gopalswamy, N. \ 2004, Geofis. Int., 43, 1

\bibitem[Lara et al.(2006)]{2006JGRA..111.6107L} Lara, A., Gopalswamy, N., Xie, H., et al.\ 2006, Journal of Geophysical Research (Space Physics), 111, A06107 

\bibitem[Liu et al.(2012)]{2012ApJ...746L..15L} Liu, Y.~D., Luhmann, J.~G., M{\"o}stl, C., et al.\ 2012, ApJl, 746, L15 

\bibitem[Liu et al.(2013)]{2013ApJ...769...45L} Liu, Y.~D., Luhmann, J.~G., Lugaz, N., et al.\ 2013, ApJ, 769, 45 

\bibitem[Lugaz et al.(2008)]{Lugaz2008} Lugaz, N., Vourlidas, A., Roussev, I.~I., et al. \ 2008, ApJl, 684, L111 

\bibitem[Lugaz et al.(2012)]{Lugaz2012} Lugaz, N., Farrugia, C.~J., Davies, J.~A., Roussev, I.~I., \& Temmer, M. \ 2012, ApJ, 759, 68 

\bibitem[Lugaz et al.(2013)]{2013ApJ...778...20L} Lugaz, N., Farrugia, C.~J., Manchester, W.~B., IV, \& Schwadron, N.\ 2013, ApJ, 778, 20 

\bibitem[Manchester et al.(2014)]{2014JGRA..119.5449M} Manchester, W.~B., Kozyra, J.~U., Lepri, S.~T., \& Lavraud, B.\ 2014, Journal of Geophysical Research (Space Physics), 119, 5449 

\bibitem[Mihalas(1978)]{1978stat.book.....M} Mihalas, D.\ 1978, San Francisco, W.~H.~Freeman and Co., 1978.~650 p.,  

\bibitem[Mishra et al.(2015)]{2015SoPh..290..527M} Mishra, W., Srivastava, N., \& Chakrabarty, D.\ 2015, Sol. Phys., 290, 527 

\bibitem[Niembro et al.(2015)]{2015ApJ...811...69N} Niembro, T., Cant{\'o}, J., Lara, A., \& Gonz{\'a}lez, R.~F.\ 2015, ApJ, 811, 69 

\bibitem[Owens et al.(2005)]{2005JGRA..110.1105O} Owens, M.~J., Cargill, P.~J., Pagel, C., Siscoe, G.~L., \& Crooker, N.~U.\ 2005, Journal of Geophysical Research (Space Physics), 110, A01105 

\bibitem[Pikel'Ner(1968)]{1968eaun.book..104P} Pikel'Ner, S.~B.\ 1968, The Earth in the Universe, 104 

\bibitem[Pizzo(1985)]{1985GMS....35...51P} Pizzo, V.~J.\ 1985, Washington DC American Geophysical Union Geophysical Monograph Series, 35, 51

\bibitem[Raga et al.(1990)]{1990ApJ...360..612R} Raga, A.~C., Binette, L., \& Canto, J.\ 1990, ApJ, 360, 612 

\bibitem[Raga et al.(2000)]{2000RMxAA..36...67R} Raga, A.~C., Navarro-Gonz{\'a}lez, R., \& Villagr{\'a}n-Muniz, M.\ 2000, RMxAA, 36, 67 

\bibitem[Riley et al.(2004)]{2004AGUFMSH24A..07R} Riley, P., Linker, J.~A., Mikic, Z., et al.\ 2004, AGU Fall Meeting Abstracts,  

\bibitem[Schwenn et al.(2004)]{2004cosp...35.2634S} Schwenn, R., dal Lago, A., Huttunen, E., \& Gonzalez, W.\ 2004, 35th COSPAR Scientific Assembly, 35, 2634 

\bibitem[Shen et al.(2011)]{2011JGRA..116.9103S} Shen, F., Feng, X.~S., Wang, Y., et al.\ 2011, Journal of Geophysical Research (Space Physics), 116, A09103 

\bibitem[Shen et al.(2013)]{2013GeoRL..40.1457S} Shen, F., Shen, C., Wang, Y., Feng, X., \& Xiang, C.\ 2013, GRL, 40, 1457 

\bibitem[Siscoe \& Odstrcil(2008)]{2008JGRA..113.0B07S} Siscoe, G., \& Odstrcil, D.\ 2008, Journal of Geophysical Research (Space Physics), 113, A00B07 

\bibitem[Stewart(1974)]{1974A&A....34..463S} Stewart, P.\ 1974, A\&A, 34, 463 

\bibitem[Temmer et al.(2011)]{2011ApJ...743..101T} Temmer, M., Rollett, T., M{\"o}stl, C., et al.\ 2011, ApJ, 743, 101 

\bibitem[Temmer et al.(2012)]{Temmer2012} Temmer, M., Vr{\v s}nak, B., Rollett, T., et al.\ 2012, ApJ, 749, 57 

\bibitem[Van Leer(1982)]{VanLeer1982} Van Leer, B., 1982, ICASE Rep. 82-30, NASA, Washington

\bibitem[Vourlidas et al.(2010)]{2010ApJ...722.1522V} Vourlidas, A., Howard, R.~A., Esfandiari, E., et al.\ 2010, ApJ, 722, 1522-1538 

\bibitem[Vr{\v s}nak(2001)]{2001JGR...10625249V} Vr{\v s}nak, B.\ 2001, JGR, 106, 25249 

\bibitem[Vr{\v s}nak et al.(2010)]{Vrsnak2010} Vr{\v s}nak, B., {\v Z}ic, T., Falkenberg, T.~V., et al.\ 2010, A\&A, 512, AA43 

\bibitem[Vr{\v s}nak et al.(2013)]{Vrsnak2013} Vr{\v s}nak, B., {\v Z}ic, T., Vrbanec, D., et al.\ 2013, Sol. Phys., 285, 295 

\bibitem[Wood et al.(2002)]{Wood2002} Wood, B.~E., M{\"u}ller, H.-R., Zank, G.~P., \& Linsky, J.~L.\ 2002, ApJ, 574, 412 

\bibitem[Xie et al.(2004)]{2004JGRA..109.3109X} Xie, H., Ofman, L., \& Lawrence, G.\ 2004, Journal of Geophysical Research (Space Physics), 109, A03109 

\bibitem[Xiong et al.(2007)]{2007JGRA..11211103X} Xiong, M., Zheng, H., Wu, S.~T., Wang, Y., \& Wang, S.\ 2007, Journal of Geophysical Research (Space Physics), 112, A11103 

\bibitem[Yashiro et al.(2004)]{2004JGRA..109.7105Y} Yashiro, S., Gopalswamy, N., Michalek, G., et al.\ 2004, Journal of Geophysical Research (Space Physics), 109, A07105 

\end{thebibliography}
\end{document}